\documentclass[numbers,compress,3p,times,review]{elsarticle} 
\usepackage[dvipsnames]{xcolor}
\usepackage{natbib}
\usepackage{amsmath,amssymb,amsfonts,amsthm}

\usepackage{pifont}
\usepackage{tabularx}
\usepackage{algorithm,algpseudocode}
\algnewcommand\algorithmicinit{\textbf{Init:}}
\algnewcommand\Init{\item[\algorithmicinit]}
\algnewcommand\algorithmicinput{\textbf{Require:}}
\algnewcommand\Input{\item[\algorithmicinput]}
\algnewcommand\myalgorithmicreturn{\textbf{Return:}}
\algnewcommand\MyReturn{\item[\myalgorithmicreturn]}
\usepackage[flushleft]{threeparttable}
\usepackage{booktabs}
\usepackage{multirow}
\usepackage{graphicx}
\usepackage{caption}
\usepackage{subcaption}
\usepackage{textcomp}
\usepackage{xcolor}
\usepackage{array}
\usepackage{dsfont}
\usepackage{commath}
\usepackage[hidelinks]{hyperref}
\usepackage[capitalise,nameinlink]{cleveref}
\usepackage{placeins}
\usepackage{import}
\crefname{equation}{}{}

\newcommand{\R}{\mathbb{R}}

\usepackage{siunitx}
\usepackage{etoolbox}

\crefname{table}{Table}{Tables}
\Crefname{table}{Table}{Tables}

\usepackage[cmintegrals,cmbraces,varbb]{newtxmath} 
\usepackage[cal=cm,bb=ams]{mathalpha} 
\renewcommand{\ttfamily}{\fontencoding{T1}\fontfamily{lmtt}\selectfont}

\let\allfigurefontsize\footnotesize

\definecolor{brown1611946}{RGB}{161,19,46}
\definecolor{chocolate2168224}{RGB}{216,82,24}
\definecolor{darkcyan0113188}{RGB}{0,113,188}
\definecolor{darkgray176}{RGB}{176,176,176}
\definecolor{darkorange25512714}{RGB}{255,127,14}
\definecolor{lightgray204}{RGB}{204,204,204}
\definecolor{steelblue31119180}{RGB}{31,119,180}

\colorlet{mygreen}{darkorange25512714}
\colorlet{myblue}{darkcyan0113188}

\usepackage{pgfplots}
\DeclareUnicodeCharacter{2212}{−}
\usepgfplotslibrary{groupplots,dateplot}
\usetikzlibrary{patterns,shapes.arrows}
\pgfplotsset{compat=newest}
\usetikzlibrary{decorations.markings}

\usepackage{pgfplotstable}
\usepgfplotslibrary{statistics}

\usepackage{tikz}
\usetikzlibrary{spy}
\usetikzlibrary{calc}
\usetikzlibrary{arrows.meta, positioning, shapes.geometric, backgrounds, fit}
\tikzset{
  block/.style = {rectangle, draw, align=center, minimum height=2em, minimum width=7.5em},
  startstop/.style = {ellipse, draw, minimum height=2em, minimum width=5em, align=center},
  decision/.style = {diamond, draw, aspect=2, inner sep=1pt, align=center},
  line/.style = {draw, -{Latex[width=2mm,length=2mm]}},
  stage/.style = {font=\small\bfseries}
}

\def\BibTeX{{\rm B\kern-.05em{\sc i\kern-.025em b}\kern-.08em
    T\kern-.1667em\lower.7ex\hbox{E}\kern-.125emX}}

\DeclareCaptionLabelFormat{notdot}{#1~\arabic{table}}
\title{Multi-scale closed-loop melt pool control for LPBF via policy optimization}
\journal{Additive Manufacturing}

\author[inst1]{Junan Lin\corref{cor1}}
\ead{linjun@ethz.ch}
\author[inst1]{Riccardo Zuliani\corref{cor1}\fnref{fn1}}
\ead{rzuliani@ethz.ch}
\author[inst3]{Bar{\i}\c{s} Kavas}
\ead{bkavas@ethz.ch}
\author[inst3]{Markus Bambach}
\ead{mbambach@ethz.ch}
\author[inst1]{John Lygeros}
\ead{jlygeros@ethz.ch}
\author[inst1,inst2]{Efe C. Balta\corref{cor2}}
\ead{efe.balta@inspire.ch}

\cortext[cor1]{Equal contribution}
\cortext[cor2]{Corresponding author}
\fntext[fn1]{The work of Riccardo Zuliani is supported as a part of NCCR Automation, a National Centre of Competence in Research, funded by the Swiss National Science Foundation (grant number 51NF40\_225155).}

\address[inst1]{Department of Information Technology and Electrical Engineering, Automatic Control Laboratory, ETH Zürich, Zürich 8092, Switzerland}
\address[inst2]{Control and Automation Group, inspire AG, Zürich 8005, Switzerland}
\address[inst3]{Department of Mechanical and Process Engineering,
ETH Z\"urich, Z\"urich 8092, Switzerland}

\begin{document}

\begin{abstract}
Laser powder bed fusion (LPBF) is a metal additive manufacturing process where temperature stabilization is of vital importance to avoid defects such as distortion and cracking. 
Existing control methods require manual tuning, increasing the risk of part failure when printing complex geometries. 
This paper introduces a dual-loop, data-driven control strategy to stabilize the surface temperature, ensuring robustness and near-optimal performance in the presence of disturbances. 
The proposed method integrates (i) an in-layer linear output feedback control with gains optimized through policy gradient, and (ii) a layer-to-layer feedforward control combining temperature trajectory optimization and iterative learning control. 
Simulation results show that the multi-scale controller effectively stabilizes the temperature even under significant model mismatch and measurement noise. 
Experimental results demonstrate that a simplified, hardware-constrained version of this method matches the state-of-the-art performance of in-situ data-driven methods, reducing mean tracking error by 3.4\% and mean input-constraint violation by 47.5\% relative to a Bayesian Optimization-tuned baseline. 
For this physical LPBF validation, the controller is tuned entirely offline using uncontrolled print data from a single calibration layer.
Our experiments also demonstrate a new class of high-frequency excitation dynamics that result in reduced vector head swelling, opening up new avenues of research in the additive manufacturing community.
This work marks one of the first successful applications of sim-to-real policy optimization in LPBF processes.
\end{abstract}

\maketitle

\section{Introduction}
Laser powder bed fusion (LPBF) is an additive manufacturing (AM) method that constructs three-dimensional parts by sequentially melting and solidifying layers of metal powder using a laser. 
The process can be summarized as follows. A layer of fine metal powder is spread over the previous layer (or the base plate). Then, a laser beam following a predefined trajectory scans and melts the powder, which upon solidifying fuses with the layers below. 
After each layer is scanned, the build platform is lowered, and the process repeats for the construction of the new layer. 
LPBF is widely adopted in industries like medicine, aerospace, and automotive since it can produce complex geometries with high precision, enable weight reduction by minimizing or removing assemblies, and accommodate a variety of materials \cite{Review_SLM, ian2015additive, frazier2014metal}. 
Despite the advantages of LPBF, the quality of the finished parts is sensitive to thermal variations due to part geometries and scan strategies \cite{yeung2020residual, thijs2010study, carter2014influence}. 
Such variations can lead to defects like inconsistent microstructure, distortion and even cracking if not properly controlled \cite{YAVARI2021109685, kumar2023comprehensive, papazoglou2020comprehensive}. 
In recent years, several control strategies across multiple time and length scales have been proposed to enable temperature stabilization in the LPBF process. 
These strategies can roughly be divided into in-layer feedback control and layer-to-layer feedforward control approaches.

In-layer feedback control adjusts laser parameters while printing a single layer to influence the melt pool and enable real-time correction of temperature errors. 
The in-layer control concept was first explored in \cite{benda1994temperature}, where on-axis optical sensing was used for melt pool regulation via power adjustments. 
Given the need for ultra-fast sensing, actuation, and computing, early work employed basic Proportional-Integral-Derivative (PID) controllers \cite{kruth2007line,kruth2007feedback}, focusing on qualitative improvements in photodiode signal stability. 
More recently, \cite{craeghs2010feedback} implemented a PI controller to address excessive energy input and overhang instability, \cite{shkoruta2022real} used a PI architecture with high-speed imaging exhibiting improved melt pool stability, \cite{kavas2024situ} employed Bayesian Optimization to auto-tune a PI controller to address overheating, \cite{renken2018model} and \cite{Renken2019} used a P-only controller to mitigate thermal accumulation effects across various geometries, while PD control was used in \cite{Wang2023} for single-layer experiments.
A full PID controller was first used in \cite{hussain2021feedback} in a simulated environment, demonstrating effective stabilization. 
More recent approaches include linear quadratic regulators (LQR) \cite{Multi-layer}, lightweight model predictive control (MPC) \cite{GP, zuliani2022batch}, small-scale supervised machine learning algorithms \cite{CARTER2024} and reinforcement learning algorithms \cite{vagenas2024multi}. 
In all cases, achieving the desired control performance often requires either manual tuning or large amounts of real-world data, making the process both time-consuming and expensive \cite{liao2024layer}.
Moreover, many of the cited works have been limited to simulation studies, without validation of the control architectures and tuning strategies on real systems.

Layer-to-layer feedforward control optimizes the inputs for the subsequent layer based on measurements collected after printing each layer.
To regulate applied energy in the next layer, \cite{Vasileska2020} introduces a layer-wise controller that monitors melt pool size fluctuations via a high-speed CMOS camera and adjusts the laser pulse frequency accordingly.
In a subsequent study, the authors implemented the same control framework on an overhanging bridge-type geometry to mitigate overheating effects arising from reduced heat conduction ~\cite{Vasileska2022}.
Residual heat caused by geometric variations was mitigated by modifying the energy input in subsequent layers, stabilizing intra-layer temperature fluctuations.
In ~\cite{kavas2023layer}, a layer-wise control strategy was introduced to regulate interlayer temperature to a target value by adjusting the laser power for each layer, whereas \cite{nahr2025advanced} presented a comparable optical tomography-based method, where processed thermal images were mapped onto the subsequent layer's scan vectors. 
In layer-to-layer control, a slower update rate (enabled by the recoating time) allows for more complex control objectives, more accurate models, and the inclusion of complex constraints in optimization.
This led to the development of trajectory optimization-based control \cite{liao2024layer} and iterative learning control (ILC) applications \cite{zuliani2022batch, Spector, kavas2023layer, model-free-ilc}, including the combined control of heating and cooling dynamics \cite{kavas2026_heat_cool}. Single-layer control-oriented models \cite{Spector} and multi-layer models \cite{wang2020layer} have also been proposed to analyze the layer-to-layer dynamics. 
Recent work has explored feedforward control strategies that update at a sublayer level. 
For example, \cite{KIRSCHBAUM2025104981} proposed a vector-by-vector technique to optimize the laser power by utilizing a coupled part-scale thermal model and small-scale melt pool model for temperature prediction.

While significant progress has been made in both in-layer and layer-to-layer control schemes, the integration of these two approaches remains relatively unexplored. 
To the best of our knowledge, \cite{zuliani2022batch} and \cite{deshmukh6519981hybrid} are the only papers that introduce a combined control scheme so far. 
\cite{zuliani2022batch} proposed a batch MPC framework that integrates in-layer MPC and layer-to-layer ILC to reject both non-repetitive and repetitive disturbances. 
However, this approach imposes substantial computational demands on the in-layer controller and does not consider multi-layer process dynamics. 
The recent work in \cite{deshmukh6519981hybrid} combines a model-based feedforward control, computed offline, with a layer-to-layer control, experimentally demonstrating improved geometric integrity and microstructural homogeneity. 
However, the layer-to-layer control component is a proprietary commercial module and is not the focus of controller synthesis or tuning. Moreover, the feedback action is only activated for excessive melt-pool intensity, leaving low-intensity regions only subjected to feedforward control. Consequently, important questions concerning controller tuning, reproducibility, robustness to model mismatch and measurement noise, and transferability across machines remain unresolved.
A further practical obstacle is obtaining reliable state estimates of the in-layer process, which is assumed by many studies \cite{Multi-layer, GP, zuliani2022batch}.
When only output measurements are available, our simulations show that in-layer control alone is insufficient to drive the output to the desired value under model mismatch and measurement noise. This highlights the necessity of a controller that combines in-layer and layer-to-layer control.

The contribution of this work is \emph{a computationally efficient multi-scale control strategy to stabilize the melt pool temperature by using the melt pool emission as an in-situ measurement signal}. The inner loop is an in-layer linear output feedback controller whose gains are optimized offline using a policy gradient method similar to~\cite{zuliani2025closed}, leveraging only an approximate process model with domain randomization for robust simulation-to-reality transfer.
The outer loop is a layer-to-layer controller combining temperature trajectory optimization (TTO)~\cite{liao2024layer} and ILC~\cite{kavas2023layer}, which specifies melt pool reference temperatures for the upcoming layer. 
The advantages of this control algorithm over the existing literature include:
%
\begin{itemize}
    \item \emph{Policy-gradient-optimized gains}: The in-layer controller gains are obtained using a policy gradient method in simulation with a simplified process model with minimal real-world tuning effort or print data availability.
    \item \emph{Computational efficiency}: The feedback control structure is computationally tractable and suitable for real-time operation as it solely relies on scalar output measurements and a single-layer TTO.
    \item \emph{Robustness to noise and uncertainty}: Both simulation and real-world experiments demonstrate near-optimal temperature stabilization even in the presence of large model mismatch and sensor noise.
\end{itemize}
The focus of this work is melt pool temperature control, resulting in control objectives specified on the melt pool time and length scales. However, the layer-to-layer control formulation can be easily recast in terms of layer-to-layer and inter-layer temperature control \cite{kavas2023layer}. 

The rest of the paper is structured as follows.
\cref{section:theory} derives the control-oriented multi-layer thermal model. \cref{sec:methods_control} presents the multi-scale control architecture and the associated controller design and policy optimization methods. \cref{sec:experiment_design} outlines the experiment design and \cref{sec:materials} introduces the LPBF setup for testing. Our experimental results modify certain components of the proposed control scheme to fit the available hardware for experimentation.
\cref{sec:results} shows the the numerical validation of the complete dual-loop controller, and the physical experimental results of the proposed controller with experimental modifications. \cref{Conclusion} summarizes our findings and suggests future research directions.
\cref{Tab:MathematicalNotations} in \ref{sec:DefinitionNotations} summarizes the notation used throughout the paper.

\section{Theory} \label{section:theory}
In this section we model the multi-layer thermal dynamics of LPBF processes and construct a control-oriented input-output model. 
Specifically, we first introduce a simplified in-layer heat equation in \cref{sec:THermodynamicsModeling} to define the input-output relation used for controller synthesis. The discretized multi-layer model in \cref{sec:ControlOrientedMultiLayerModel} then incorporates inter-layer heat exchange, build-plate and ambient boundary terms, and the cooling and recoating transition.
The model extends the early work in~\cite{liao2024layer} which introduced one of the first multi-layer control oriented models, adapted in recent studies such as~\cite{kavas2023layer,kavas2026_heat_cool}.

\subsection{Control-oriented In-layer Modeling}\label{sec:THermodynamicsModeling}
Following \cite{Spector, GUSAROV2007975, VERHAEGHE20096006, roy2018heat} we use a simplified partial differential equation (PDE) model that neglects highly nonlinear phenomena of the laser process such as phase transition and volumetric thermal conduction. 
While we explicitly account for vertical heat transfer between subsequent layers with the layer-to-layer model in the next section, here we assume that the heat transfer of each layer is confined to the in-plane direction and neglect any variation along the layer's height. 
We assume the printed part consists of $N$ layers, each governed by a heat equation that evolves in two spatial dimensions. 
For each layer $k = 1, 2, \dots, N$, the temperature dynamics are described by
\begin{equation}
    \frac{\partial{H_{k}(r, \tau)}}{\partial{\tau}} = \nabla(\kappa(r)\nabla T_{k}(r,\tau))+q_k(r,\tau), ~\tau\in[0,\tau_{f}]\text{,}
    \label{eqn:HeatEquation}
\end{equation}
where $r\in\R^2$ denotes the in-plane position within the two-dimensional cross-section of layer $k$, $H_k$ is the volumetric enthalpy, 
$\kappa$ is the thermal conductivity, $T_k$ is the temperature of layer $k$, and $q_k$ is the heat flux provided by the laser to layer $k$.
We use $\tau_f$ to denote the processing time, which is assumed to be constant across all layers, and includes the layer printing time $\tau_p$ and the recoating time $\tau_c$
\begin{equation}
    \tau_f=\tau_p+\tau_c.
    \label{eqn:ProcessingTime}
\end{equation}
The relationship between volumetric enthalpy $H_k$ and temperature $T_k$ for each layer is
\begin{equation}
    H_k(r,\tau)=c_{k}T_k(r,\tau),
    \label{eqn:H=c_hT}
\end{equation}
where $c_{k}$ is the volumetric heat capacity of layer $k$, assumed to be uniform across nodes within each layer.

We assume the laser exposure moves along a fixed and known trajectory $(p(\tau),z_N)$ defined by the scan strategy, 
where $p:[0,\tau_f]\to \R^2$ represents the in-plane $x$-$y$ position within the two-dimensional cross-section of layer $N$, and $z_N\in\R$ is the height of layer $N$. 
The laser directly affects only the topmost layer $N$; specifically, the heat flux $q_k(r,\tau)$ at the position $r\in\R^2$, time $\tau$, and layer $k$ is given by
\begin{equation}
    q_k(r,\tau) =  
        \begin{cases}
        \mathcal{B}(r,\tau)u(\tau) & \text{if $k = N$,}\\
        0 & \text{otherwise,}\\
    \end{cases}
    \label{eqn:HeatFlux}
\end{equation}
where $u:[0,\tau_f]\to\R$ represents the laser power, which is zero during the recoating time, and $\mathcal{B}(r,\tau)$ is the intensity of the laser beam. 
Since the laser trajectory is fixed, the only available control variable is the laser power $u(\tau)$.

The intensity of the beam within the beam radius $a$ can be represented by \cite{Spector}
\begin{equation}
    \mathcal{B}(r,\tau)= \begin{cases}
    \frac{3\alpha}{\pi a^2}\left( 1-\frac{\|r-p(\tau)\|^2}{a^2} \right)^2, & \text{if } \|r-p(\tau)\| \leq a, \\
    0 & \text{otherwise},
    \end{cases}
    \label{eqn:IntensityGaussianBeam}
\end{equation}
where $\alpha\in[0,1]$ is the fraction of the power absorbed by the material. 
The output is chosen as
\begin{equation}
    y(\tau)=\int \mathcal{B}(r,\tau)^\top T_N(r,\tau) dr,
    \label{eqn:Output}
\end{equation}
where the integral is computed on the 2D cross-section of the layer. Since the pyrometer effectively captures an average of the surface temperature under the laser spot, 
this output choice mimics the real sensor signals \cite{kavas2023layer, Spector}.

The objective is to drive the output $y$ to a desired temperature $y_d$ during the printing period of each layer while satisfying constraints on the laser power, 
as described in the following optimization problem.
\begin{equation}
\begin{aligned}
    \min_{u(\tau)} \quad & \int_{\tau=0}^{\tau_p} \norm{y(\tau) - y_d}^2 d\tau, \\
    \text{s.t.} \quad &(\ref{eqn:HeatEquation}), (\ref{eqn:H=c_hT}), (\ref{eqn:HeatFlux}), (\ref{eqn:IntensityGaussianBeam}), (\ref{eqn:Output}),\\
    & u_{\text{min}} \leq u(\tau) \leq u_{\text{max}},~\tau\in[0, \tau_p],\\
    & u(\tau) = 0, ~ \tau \in [\tau_p, \tau_f].
    \label{eqn:MainProblem}
\end{aligned}
\end{equation}

\subsection{Control-oriented Multi-layer Modeling}\label{sec:ControlOrientedMultiLayerModel}
\begin{figure}
\centering
\scalebox{0.8}{
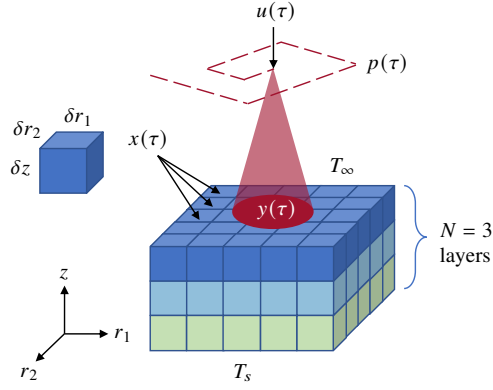
}
\caption{Multi-layer model adapted from \cite{Multi-layer}. 
The light green bottommost layer has direct contact with the thermostatic building plate, 
and the dark blue topmost layer has direct contact with the ambient atmosphere and the laser.}%
\label{fig:MultiLayerModel}
\end{figure}

We now adapt the control-oriented multi-layer model proposed in \cite{Multi-layer} to fit our control architecture. 
To this end, we spatially discretize the PDE in (\ref{eqn:HeatEquation}) on each layer into $n=n_{r_1} \times n_{r_2}$ nodes $I_{k}^i$, 
each consisting of a rectangular volume of size $\delta r_1\times \delta r_2 \times \delta z$ (compare \cref{fig:MultiLayerModel} for a visual depiction with $N=3$ layers and $n_{r_1}=n_{r_2}=5$). 
For simplicity, we assume that $\delta r_1\equiv\delta r_2=:\delta r$.

The heat transfer equation of each layer depends on the material (powder or solid metal), 
and the nature of the neighboring layers (substrate, powder layer, solid layer). 
The heat transfer coefficient between neighboring coplanar nodes is assumed to be constant throughout each layer
\begin{align}
\kappa_k =
\begin{cases}
\kappa_\text{p} & \text{if $k = N$},\\
\kappa_\text{d} & \text{otherwise},
\end{cases} \label{eq:heat_transfer_coeff}
\end{align}
where $\kappa_\text{p}$ and $\kappa_\text{d}$ are the thermal conductivities of the powder and the solidified (dense) material, respectively. 
Similarly, the heat transfer coefficients between any node $I_k^i$ and its neighbors in layers $k-1$ and $k+1$ are respectively defined as
\begin{align}
    \kappa_{k-1,k} =
    \begin{cases}
    \kappa_{\text{pd}} \! & \text{if $k = N$},\\
    \kappa_\text{d} \! & \text{otherwise},\\
    \end{cases}\label{eqn:HeatCoef_k-1_k}~~
    \kappa_{k+1,k} =
    \begin{cases}
    h_\infty \delta z \! & \text{if $k = N$},\\
    \kappa_{\text{pd}} \! & \text{if $k = N \! - \! 1$},\\
    \kappa_\text{d} \! & \text{otherwise},
    \end{cases}
\end{align}
where $\kappa_{\text{pd}}$ is the thermal conductivity between powder and solid material, 
and $h_\infty$ is the convection coefficient between nodes and the atmosphere.

The volumetric heat capacity of any node within layer $k$ is given by
\begin{equation}
\begin{aligned}
    c_k =
    \begin{cases}
    (1-\epsilon)c_\text{d} & \text{if } k = N,\\
    c_\text{d} & \text{otherwise},\\
    \end{cases}
    \label{eqn:HeatCapacity}
\end{aligned}
\end{equation}
where $c_\text{d}$ and and $(1-\epsilon)c_\text{d}$ are the volumetric heat capacities of the solidified metal and the powder, respectively, where $\epsilon$ is the porosity of the metal.

For each node $I_k^i$, we denote the set of neighboring nodes in layer $k$ with $\mathcal{N}_{k}^i$, 
and the set of neighboring nodes in layers $k+1$ and $k-1$ with $\mathcal{N}_{k,k+1}^i$ and $\mathcal{N}_{k,k-1}^i$, respectively. 
Note that the sets $\mathcal{N}_{k,k+1}^i$ and $\mathcal{N}_{k,k-1}^i$ can be empty if $k=N$ or $k=0$. 
The heat equation for node $i$ in layer $k$ is then given by
\begin{equation}
\begin{aligned}
    \delta r^2 \delta z \cdot c_{k} \dot{T}_k^{i}(\tau) = & -{\sum_{j\in\mathcal{N}_{k}^i}} \delta z \cdot \kappa_k (T_k^{i}(\tau)-T_k^{j}(\tau)) - {\sum_{j\in\mathcal{N}_{k,k-1}^i}} \frac{\delta r^2}{\delta z} \cdot \kappa_{k-1,k} (T_k^{i}(\tau)-T_{k-1}^{j}(\tau)) \\
    & - {\sum_{j\in\mathcal{N}_{k,k+1}^i}} \frac{\delta r^2}{\delta z} \cdot \kappa_{k+1,k} (T_k^{i}(\tau)-T_{k+1}^{j}(\tau)) + \sigma_k b_k^{i}(\tau) u(\tau)\text{,}
    \label{eqn:TemperatureSingleNode}
\end{aligned}
\end{equation}
where $b_k^{i}(\tau)$ is the beam intensity in (\ref{eqn:IntensityGaussianBeam}) evaluated at time $\tau$ at the center of corresponding node, 
and $\sigma_k$ represents the top-most layer, which is affected by the laser beam; thus, $\sigma_k=1$ when $k=N$ and $\sigma_k^{i}=0$ otherwise.
The temperature $T_{N+1}^i$ is constant for all nodes $i$ and equals the ambient temperature $T_\infty$. 
Similarly, $T_0^i$ equals the temperature of the build plate $T_\text{s}$ for all $i$.

To write (\ref{eqn:TemperatureSingleNode}) in a more compact form, let $x_k=[T_k^{1},...,T_k^{n}]^\top$ denote the state for all nodes in layer $k$, 
and $B(\tau)=[b_N^{N}(\tau), ..., b_N^{n}(\tau)]^\top$ denote the beam intensity for all nodes in the topmost layer $N$. 
We define $\mathcal{L}=D-W$ as the common graph Laplacian for the graph of each layer \cite{mesbahi2010graph}, where $D$ is the diagonal degree matrix and $W$ is the adjacency matrix.
With this notation, the heat equation of layer $k$ is given by
\begin{align}
    \delta r^2 \delta z \cdot c_k \dot{x}_{k}(\tau) = & - \delta z \cdot \kappa_k \mathcal{L} x_{k}(\tau) + \frac{\delta r^2}{\delta z} \cdot \kappa_{k-1,k}(x_{k-1}(\tau)-x_{k}(\tau)) \notag \\
    & + \frac{\delta r^2}{\delta z} \cdot \kappa_{k+1, k} (x_{k+1}(\tau)-x_{k}(\tau)) + \sigma_k B(\tau)u_N(\tau),
    \label{eqn:HeatEquationAll}
\end{align}
where $x_{N+1}(\tau) \equiv T_\infty \mathbf{1}_n$, $x_{0}(\tau) \equiv T_s \mathbf{1}_n$, with $\mathbf{1}_n\in\R^n$ being the vector of all ones. 
The output in (\ref{eqn:Output}) becomes
\begin{equation}
    y_N(\tau)=B(\tau)^\top x_N(\tau).
    \label{eqn:OutputDiscretizedSpatially}
\end{equation}

Defining $X_N=[x_1^\top,x_2^\top,...,x_N^\top]^\top$ as the state of the combined multi-layer dynamics, 
the heat equation in (\ref{eqn:HeatEquationAll}) can be compactly represented as a linear time-varying (LTV) system in the standard form
\begin{align}
    \begin{split}
    \dot{X}_N(\tau) &= A_N X_N(\tau) + B_N(\tau)u_N(\tau) + d_N\text{,}\\
    y_N(\tau) &= B_N(\tau)^\top X_N(\tau)\text{,}\\
    \end{split}
    \label{eqn:LTVContinuous}
\end{align}
where $\tau\in[0,\tau_f]$, and
\begin{align*}
B_N(\tau) & = \begin{bmatrix} 0 & 0 & 0 & \dots & 0 & \frac{1}{\delta r^2 \delta z c_N} B(\tau) \end{bmatrix}^\top,\\
d_N & = \begin{bmatrix} \frac{T_s}{\delta z^2 c_1}\mathbf{1}_n & 0 & 0 & \dots & 0 & \frac{T_\infty}{\delta z^2 c_N}\mathbf{1}_n \end{bmatrix}^\top,\\
A_N & = \begin{bmatrix}
A_{N,11} & \frac{\kappa_{2,1}}{\delta z^2 c_1} I_n & 0 &\cdots & 0 \\
\frac{\kappa_{1,2}}{\delta z^2 c_2} I_n & A_{N,22} & \frac{\kappa_{3,2}}{\delta z^2 c_2} I_n & \cdots & 0 \\
0 & \frac{\kappa_{2,3}}{\delta z^2 c_3} I_n & A_{N,33} & \cdots & 0 \\
\vdots & \vdots & \vdots & \ddots & \vdots \\
0 & 0 & 0 & \cdots & \frac{\kappa_{N,N-1}}{\delta z^2 c_{N-1}} I_n\\
0 & 0 & 0 & \cdots & A_{N,NN}\\
\end{bmatrix},
\end{align*}
with
\begin{align*}
A_{N,ii} = \begin{cases}
-\frac{\kappa_i}{\delta r^2 c_i} \mathcal{L} - \frac{\kappa_{i-1,i}+\kappa_{i+1,i}}{\delta z^2 c_i} I_n & \text{if } i\in \mathbb{Z}_{[1,N-1]},\\
-\frac{\kappa_N}{\delta r^2 c_N} \mathcal{L} - \frac{\kappa_{N-1,N}}{\delta z^2 c_N} I_n & \text{if } i=N.
\end{cases}
\end{align*}
The output dynamics in~\eqref{eqn:LTVContinuous} is the adjoint of the input matrix to simplify the dynamics and the computational overhead. The output matrix can be chosen with independent parameters and structure depending on the application, which we do not study in this work.

After layer $N$ is printed, the build plate is lowered to accommodate a new layer indexed by $N+1$, 
thereby enlarging the build volume. A fresh layer of powder is then spread via the recoating process. 
During this period, the build volume cools, and heat flows into the new powder layer. 
This process is modeled by the \emph{next-layer operator} $\mathcal{S}$, defined as
\begin{equation}
\begin{aligned}
    X_{N+1}(0) = \mathcal{S}\Bigl(X_{N}(\tau_p),\tau_c\Bigl) =
    \begin{bmatrix}
    e^{A_N \tau_c} X_{N}(\tau_p) + A_N^{-1} (e^{A_N \tau_c} - I) d_N \\
    T_s\mathbf{1}_n
    \end{bmatrix}\text{,}
    \label{eqn:NewLayerDetailed}
\end{aligned}
\end{equation}
where $X_{N+1}(0)$ is the state at the start of the print of layer $N+1$. 
The introduction of the next-layer operator indicates that $A_N$, $B_N(\tau)$ and $d_N$ should be changed after the cooling of each topmost layer, due to the enlargement of the state.

The linear ODE in (\ref{eqn:LTVContinuous}) can be discretized in time to obtain
\begin{equation}
    \begin{cases}
        X_N[t+1] = \Bar{A}_{N} X_N[t] + \Bar{B}_{N}[t] u_N[t] + \Bar{d}_{N}\text{,}\\
        y_N[t] = \Bar{C}_{N}[t] X_N[t]\text{,}\\
    \end{cases}
    \label{eqn:LTVDiscrete}
\end{equation}
where $t$ is the discrete-time index, $\bar{A}_N=e^{A_N \Delta t}$, $\bar{B}_N[t] = A_N^{-1} (\Bar{A}_N-I) B_N(t\Delta t)$, 
$\Bar{d}_N=A_N^{-1} (\Bar{A}_N-I) d_N$, $\Bar{C}_N[l]=B_N(t\Delta t)^\top$, and $\Delta t$ is the sampling period.
The discretization is not exact due to the choice of $\bar{B}_N[t]$; 
however, the discretization error vanishes as $\Delta t \to 0$, ensuring that \cref{eqn:LTVDiscrete} becomes accurate for sufficiently small $\Delta t$. 
This is because $B_N(\tau)$ is continuous, and thus does not change significantly when the sampling instants are sufficiently close. 
This approximation offers a substantial computational advantage compared to exact discretization, as it avoids numerical integration, 
which can become prohibitive when the number of layers (or nodes) is very large.

The model in (\ref{eqn:LTVDiscrete}) can be written equivalently as a lifted system, where $t_p = \tau_p / \Delta t$ is the time step when the printing process of a layer terminates. 
Denoting with $y_N = [y_N[1],...,y_N[t_p]]^\top$ and $u_N=[u_N[0],...,u_N[t_p-1]]^\top$ the output and input sequences for layer $N$, the lifted system output is given by
\begin{equation}
    y_N = Y_{N}^u u_N + Y_{N}^x X_N[0] + Y_{N}^d \bar{d}_N\text{,}
    \label{eqn:LiftedSystem}
\end{equation}
where $Y_N^u$ is the lower triangular matrix whose $ij$-th entry is
\begin{align*}
Y_{N,ij}^u= \begin{cases}    
\bar{C}_N[i+1]\bar{A}_N^{i-j}\bar{B}_N[j] & \text{if } i \geq j, \\
0 & \text{otherwise},
\end{cases}
\end{align*}
and
\begin{align*}
Y_{N}^x & = \operatorname*{col}(\bar{C}_N[1] \bar{A}_N, \bar{C}_N[2] \bar{A}_N^2, \dots, \bar{C}_N[t_p] \bar{A}_N^{t_p}),\\
Y_{N}^d & = \operatorname*{col}(\bar{C}_N[1], \bar{C}_N[2] (\bar{A}_N + I), \bar{C}_N[t_p] {\textstyle \sum_{i=0}^{t_p-1}} \bar{A}_N^{i}).
\end{align*}

To account for model uncertainty, we let the matrices $Y_N^u$, $Y_N^x$, and $Y_N^d$ in (\ref{eqn:LiftedSystem}) depend on an unknown \emph{true} parameter $\theta$, 
and let $\theta'$ denote the (known) estimated value of $\theta$. 
We further assume that the true parameter is contained within an interval of possible parameters $[\theta_\text{min},\theta_\text{max}]$. 
We use $Y_N^u(\theta)$, $Y_N^x(\theta)$, $Y_N^d(\theta)$ to denote the real values of the matrices in \cref{eqn:LiftedSystem} and $Y_N^u(\theta')$, 
$Y_N^x(\theta')$, $Y_N^d(\theta')$ to denote the estimated values.


\section{Multi-Scale Control System Design}
\label{sec:methods_control}
\subsection{Control Architecture}

\begin{figure}[b]
\centering
\scalebox{0.8}{
    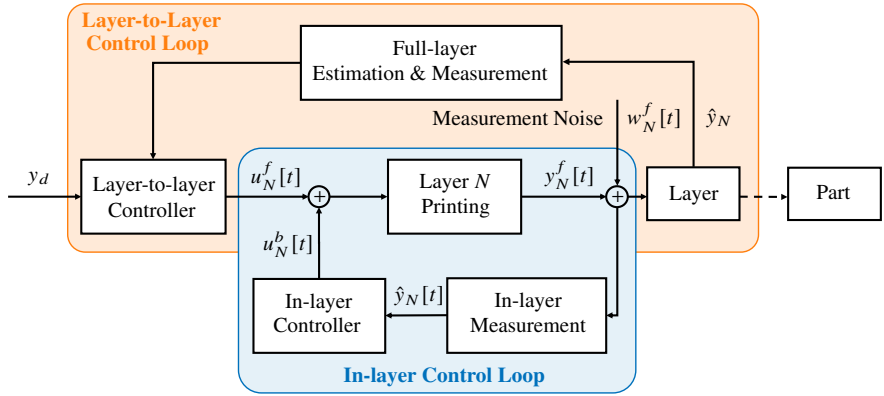
}
\caption{The proposed multi-scale LPBF feedback control architecture consisting of a dual-loop architecture.}
\label{fig:SLMControlArchitecture}
\end{figure}

\begin{figure}[t]
\centering
\begin{tikzpicture}[font=\footnotesize,node distance=0.4cm and 0.5cm,semithick]

\node[startstop] (start) {Start};
\node[block, below=of start] (determineY) {Determine $Y_N^u(\theta'), Y_N^x(\theta')$\\ and $Y_N^d(\theta')$.};
\node[block, below=of determineY] (y1u1) {Solve \eqref{eqn:LayerToLayerOptimization:first_layer} to obtain $u_1^f, y_1$.};
\node[block, below=of y1u1] (trainK) {Solve \cref{eqn:InLayerOptimization} to obtain the \\ control gain $K$};

\node[block, right=1.95cm of start] (initN) {Initialize layer $N=1$};
\node[block, below=0.85cm of initN] (runFeedForward) {Solve \cref{eqn:LayerToLayerOptimizationSimplified} to obtain\\ feedforward input $u_N^f[t]$ \\ for layer $N$};
\node[block, below=0.85cm of runFeedForward] (initT) {Initialize time index $t=0$};
\node[block, below=0.85cm of initT] (deriveFeedback) {Derive feedback input $u_N^b[t]$\\ with gain $K$};

\node[block, left=0.65cm of deriveFeedback] (print) {Print for time $t$ in layer $N$\\ with $u_N^f[t]$ and $\sum_{k=1}^N u_k^b[t]$,\\ set $t \leftarrow t+1$};
\node[decision, below=of print] (checkT) {$t \geq t_p$?};
\node[decision, below=of checkT] (checkN) {$N \geq N_{\text{max}}$?};
\node[block, right=2cm of checkN] (incN) {Let $N \leftarrow N + 1$};
\node[coordinate, left=of deriveFeedback] (leftloop) {};

\node[startstop, below=of checkN] (end) {end};

\draw[-latex] (start) -- (determineY);
\draw[-latex] (determineY) -- (y1u1);
\draw[-latex] (y1u1) -- (trainK);
\draw[-latex] (trainK.east) -- ++(0.75,0) |- (initN.west);

\draw[-latex] (initN) -- (runFeedForward);
\draw[-latex] (runFeedForward) -- (initT);
\draw[-latex] (initT) -- (deriveFeedback);
\draw[-latex] (deriveFeedback) -- (print);
\draw[-latex] (print.south) -- (checkT);

\draw[-latex] (checkT) -- node[right] {Yes} (checkN);
\draw[-latex] (checkN) -- node[right] {Yes} (end);
\draw[-latex] (checkN) -- node[above] {No} (incN);
\draw[-latex] (incN.east) -- ++(0.6cm,0cm) -- ([xshift=0.45cm]runFeedForward.east) -- (runFeedForward.east);

\draw[-latex] (checkT.east) -- node[pos=0.335, above] {No} ([yshift=-0.95cm]deriveFeedback.south) -- (deriveFeedback.south);

\begin{pgfonlayer}{background}
    \node[draw=myblue, thick, dashed, fill=myblue!10, fit=(start) (determineY) (trainK), inner sep=0.2cm] (offlinebg) {};

    \node[anchor=north west, thick, text=myblue, font=\bfseries, xshift=0.4cm, yshift=0.6cm] at (offlinebg.north west) {Offline design phase};

    \node[anchor=north west, thick, text=mygreen, font=\bfseries, xshift=5.3cm, yshift=0.6cm] at (offlinebg.north west) {Online phase};

    \filldraw[draw=mygreen, dashed, fill=mygreen!10, thick]
        ([xshift=-0.5cm,yshift=0.2cm]initN.north west) --
        ([xshift=0.9cm,yshift=0.2cm]initN.north east) --
        ([xshift=0.9cm,yshift=-8.9cm]initN.north east) --
        ([xshift=-4.45cm,yshift=-8.9cm]initN.north east) --
        ++(0,-1.35cm) -- ++(-2.35cm,0) -- ++ (0,3.875cm) --
        ++ (-0.7cm,0) -- ++(0,1.48cm) -- 
        ([xshift=-0.5cm,yshift=-4.9cm]initN.north west) --
        cycle;
\end{pgfonlayer}

\end{tikzpicture}
\caption{Procedure diagram for training and deploying the dual-loop controller. Note that $N_\text{max}$ is the desired number of layers, and $t_p$ is the printing time for a single layer.}\label{fig:AlgorithmDiagram}
\end{figure}
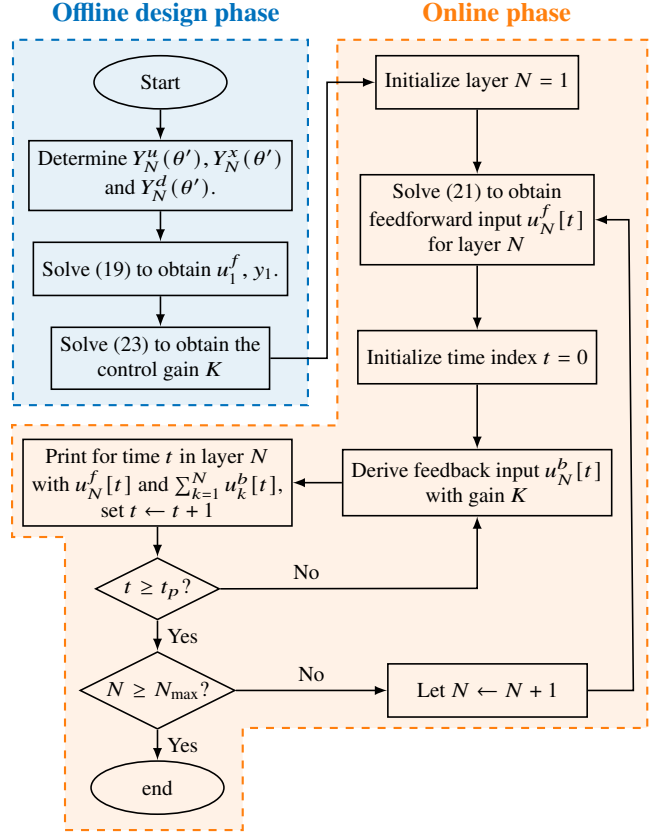

The proposed control architecture comprises two nested control loops, illustrated in \cref{fig:SLMControlArchitecture}. 
A more detailed description of the controller is provided in \cref{fig:AlgorithmDiagram}, dividing it into offline and online operations. 
The offline phase determines the best estimate of the process model based on information about materials, part geometry, and laser path. It also determines the gains used by the in-layer controller by solving an optimization problem that aims to robustify against noise and uncertainty in the model estimate. 
The online phase is where the build takes place. In this phase, a feedforward control input is combined with the feedback in-layer controller trained in the offline phase. The feedforward input is updated across layers using output measurements collected from the system exploiting the process model.

The \emph{layer-to-layer control loop} provides optimized feedforward input profiles for an entire layer, 
while the \emph{in-layer control loop} uses real-time pyrometer measurements to adjust the laser power at run-time as the layer is being printed. 
The full input is
\begin{equation}
    u_{N}[t] = \Pi_{U} \Bigl(u^f_N[t] + \sum_{k=1}^N u_k^b[t] \Bigl)\text{,} \\
    \label{eqn:FullInput}
\end{equation}
where $u_N^f[t]$ is the feedforward layer-to-layer control for layer $N$, $u_{k}^b[t]$ is the feedback in-layer control for layer $k$, 
$\Pi_U$ denotes the projection onto the set $U=[u_\text{min}, u_\text{max}]$, 
and $u_{\text{min}}, u_{\text{max}}$ are the minimal and maximal laser power respectively.

The use of a nested control architecture is motivated by the layer-to-layer nature of the LPBF process. 
The outer loop utilizes layer-wide measurements to iteratively correct the feedforward input trajectory, 
ensuring that the tracking performance improves layer after layer. 
However, this controller alone does not have any immediate effects within each layer. 
This motivates the second in-layer control loop, which is able to quickly reduce the tracking error by utilizing online measurements.
We next explain in detail both the control loops, starting with the layer-to-layer part.

\subsection{Layer-to-layer Control Loop Design}\label{sec:LayerToLayerControlLoopDesign}

The layer-to-layer control loop generates feedforward input trajectories $u^f_N$ using measurements $\hat{y}_N$ collected after printing each layer. 
The controller is formulated as a convex optimization problem combining TTO and ILC. 
To maintain computational tractability, we only consider single layer dynamics
\begin{align*}
y_N = Y^u(\theta) u_N^f + Y^x(\theta) X[0] + Y^d(\theta) \bar{d}(\theta).
\end{align*}
as derived in (\ref{eqn:LiftedSystem}), where we omitted the subscript $1$ for simplicity by setting $Y^u(\theta):=Y_1^u(\theta)$, $Y^x(\theta):=Y_1^x(\theta)$, $Y^d(\theta):=Y_1^d(\theta)$, and $\bar{d}(\theta):=\bar{d}_1(\theta)$.
These matrices and vectors depend on the parameter $\theta$ which denotes physical quantities of the LPBF process (such as laser power absorption, porosity). 
The feedforward input for the first layer can be obtained without any measurement from the system by leveraging knowledge of the model dynamics
\begin{align}
\begin{split}
\operatorname*{minimize}_{u_1^f, y_1} \quad & J_1(u_1^f, y_1) = \sum_{j=1}^{t_p} \|{y_1[j] - y_d}\|_{Q}^2 + \|{u_1^f[j-1]}\|_{R}^2 \\
\text{subject to} \quad &y_1 = Y^u(\theta') u_1^f + Y^x(\theta') X[0] + Y^d(\theta') \bar{d}(\theta'),\\
&X[0] = T_s\mathbf{1}_n,\\
&u_{\text{min}} \leq u_1^f[t] \leq u_{\text{max}},~t=0,1,...,t_p-1.
\end{split}\label{eqn:LayerToLayerOptimization:first_layer}
\end{align}
where $u_1^f$ and $y_1$ are the feedforward input and the nominal output for the first layer, respectively, and $y_d$ is the desired output.
The parameter $\theta'$ is the best available estimate of the model parameters, obtained during the offline phase of our algorithm, which we assume to be constant for all layers.
After completing the first layer and obtaining the true output measurement $\hat{y}_1$, the optimization problem we solve for each layer becomes
\begin{align}
\begin{split}
\operatorname*{minimize}_{u_N^f, y_N} \quad &J_N(u_N^f, y_N) = \sum_{j=1}^{t_p} \|{y_N[j] - y_d}\|_{Q}^2 + \|{u_N^f[j-1]}\|_{R}^2 \\
\text{subject to} \quad &y_N = Y^u(\theta') u_N^f + Y^x(\theta') X[0] + Y^d(\theta') \bar{d}(\theta') + \sum_{k=1}^{N-1} L_y (\hat{y}_{k} - y_{k}),\\
&X[0] = T_s\mathbf{1}_n,\\
&u_{\text{min}} \leq u_N^f[t] \leq u_{\text{max}},~t=0,1,...,t_p-1,
\end{split}\label{eqn:LayerToLayerOptimization}
\end{align}
where $L_y\in(0,2)$ is a learning gain, $u_N^f$ and $y_N$ are the feedforward input and the nominal output for layer $N$, respectively, and $\hat{y}_k$ is measured output for layer $k$. 
The quadratic cost in \eqref{eqn:LayerToLayerOptimization:first_layer} and \eqref{eqn:LayerToLayerOptimization} mimics the standard linear-quadratic regulator (LQR) cost. It penalizes the output tracking error $y_N[j]-y_d$, as well as the control input, where the positive semi-definite matrices $Q$ and $R$ can be tuned to reduce the aggressiveness of the control update (by increasing $R$) or to increase the tracking performance (by increasing $Q$).

The term $\sum_{k=1}^{N-1} L_y (\hat{y}_{k} - y_{k})$ is used to iteratively correct the error caused by model reduction, incorrect parameter estimation, and measurement noise, using measurement of outputs collected in previous layers.
This term, which acts as a form of layer-to-layer integral controller, is similar to the well-known \emph{norm-optimal ILC} \cite{amann1996optimal, NO-ILCwithConstraint}, which is retrieved by letting $L_y=1$.
Here, we let $L_y$ be a design variable for additional flexibility.
Observe that this term is absent in \eqref{eqn:LayerToLayerOptimization:first_layer} when $N=1$.

By imposing $X_1[0]=T_s \mathbf{1}_n$, we assume that the printed layers reach their equilibrium temperature before the print of the next layer begins. This amounts to an adaptive waiting time between layers, which was recently demonstrated to improve the overall inter-layer temperature response in~\cite{kavas2026_heat_cool}.
Alternatively, the initial temperature could be set to the temperature of the previous layer at the start of the new layer, $X_{N-1}[0]$. This modification recovers the layer-to-layer controller formulation in our previous work~\cite{kavas2023layer}, which shows robust closed-loop performance in experiments. Overall, the layer-to-layer formulation~\eqref{eqn:LayerToLayerOptimization} generalizes existing controllers in the literature to provide a unified framework.

Problem \eqref{eqn:LayerToLayerOptimization} is a convex quadratic program (QP) that can be solved efficiently using readily available software packages. 
Choosing $z=u_N^f$, we can express \eqref{eqn:LayerToLayerOptimization} in standard form as
\begin{align}
\begin{split}
\operatorname*{minimize}_{z} \quad &\frac{1}{2} z^\top H z + f^\top z \\
\text{subject to} \quad &Gz \leq S,\\
\end{split}\label{eqn:LayerToLayerOptimizationSimplified}
\end{align}
with $H = 2 (\bar{R} + Y^u (\theta')^\top \bar{Q} Y^u (\theta')$), $f = 2 (\phi - \bar{y}_d)^\top \bar{Q} Y^u (\theta')$, 
$\phi = Y^x (\theta') X[0] + Y^d (\theta') \bar{d} (\theta') + \sum_{k=1}^{N-1} L_y (\hat{y}_k - y_k)$, 
$G = \operatorname*{col}[I_n, -I_n]$, and $S = \operatorname*{col}[\bar{u}_{\text{max}}, -\bar{u}_{\text{min}}]$, 
where $\bar{Q}$, $\bar{R}$ are block diagonal matrices and $\bar{y}_d$, $\bar{u}_\text{max}$, 
$\bar{u}_\text{min}$ are vectors obtained by stacking repeated copies of $Q$, $R$, $y_d$, $u_\text{max}$, and $u_\text{min}$, respectively.

\subsection{In-layer Control Loop Design}\label{sec:InLayerControlLoopDesign}

The in-layer control loop uses real-time output measurement $\hat{y}_N[t]$, collected as the layer is being printed, to adjust the laser power, enabling real-time feedback. 
The feedback term $u_N^b[t]$ is given by the following linear output disturbance feedback scheme
\begin{equation}
    u_{N}^b[t]=\sum_{i=0}^{t} K_{t}[i] e_{N}[i]\text{,}
    \label{eqn:DisturbanceFeedback}
\end{equation}
where $K_{t}[i]$ is a feedback gain and $e_{N}[i] = y_{N}[i] - \hat{y}_N[i]$. 
The gains $K_t[i]$ must be determined offline before starting the print of the first layer. 
To obtain them, we solve the following optimization problem, where $K$ is the matrix with $K_t[i]$ on its $t$-th row and $i$-th column:
\begin{equation}
\begin{aligned}
    \operatorname*{minimize}_K \quad &J_{\text{in}}(K) = \mathbb{E}_{w,\theta} \Bigl[~\frac{1}{t_p} \sum_{j=1}^{t_p} \norm{\hat{y}[j] - y_d}^2~\Bigl] \\
    \text{subject to} \quad &y = Y^u(\theta) u + Y^x(\theta) X[0] + Y^d(\theta) \bar{d}(\theta),\\
    &\hat{y}[t] = y[t] + w[t], \\
    &w[t] \sim \text{Uniform}[w_{\text{min}}, w_{\text{max}}],\\
    &\theta \sim \text{Uniform}[\theta_{\text{min}}, \theta_{\text{max}}],\\
    &u[t] = \Pi_U \Bigl(u^f[t]+u^b[t] \Bigl),\\
    &u^b[t]=\sum_{i=0}^{t} K_{t}[i] e[i], ~t=0,1,...,t_p-1,\\
    &e[i] = y[i] - \hat{y}[i],~i=0,1,\dots,t,\\
    &X[0] = T_s\mathbf{1}_n.
    \label{eqn:InLayerOptimization}
\end{aligned}
\end{equation}

In \cref{eqn:InLayerOptimization}, the goal is to choose a value of $K$ that ensures the best output tracking (by minimizing $\|y[j]-y_d\|^2$) in the presence of noise $w[t]$ and model uncertainty (on $\theta$). 
By explicitly accounting for model uncertainty and noise, we ensure that any minimizer $K$ of the problem is robust when deployed in closed-loop operation.

In \cref{eqn:InLayerOptimization}, $\hat{y}, y$ denote the measured output and the nominal output, respectively, 
$w_{\text{min}}$ and $w_{\text{max}}$ denote estimated bounds on the measurement noise, $\theta_{\text{min}}, \theta_{\text{max}}$ are estimates of the minimum and maximum parameter values, 
$u^f$ is the feedforward input obtained by solving the layer-to-layer problem \cref{eqn:LayerToLayerOptimization}, and $\bar{X}[0]$ is the initial state of the system.  
The framework can be extended to robustify against variations of the initial temperature, which is an interesting avenue for future work. 
Here, we focus on model uncertainty and measurement noise as they have high uncertainty and influence on the closed-loop dynamics.

To solve \cref{eqn:InLayerOptimization} efficiently, we apply a gradient-based scheme to an unconstrained reformulation of \cref{eqn:InLayerOptimization} using penalty functions for the input constraints. 
For a given value of $w$ and $\theta$, the gradient of the loss function $J_{\text{in}}(K)$ is given by
\begin{equation}
    \nabla J_{\text{in}}(K) = \frac{\partial J_{\text{in}}(K)}{\partial y} \frac{\partial y}{\partial u} \frac{\partial u}{\partial u^b} \frac{\partial u^b}{\partial K},
    \label{eqn:FullGradient}
\end{equation}
where $\partial J_{\text{in}}(K)/\partial y = 2 (\hat{y}-y_d)^\top/ t_p $, $\partial y/\partial u = Y_{1}^u(\theta)$, $\partial u/\partial u^b = \text{diag} ( \text{sat}(u^f+u^b) )$, with $\operatorname*{sat}(u)=1$ if $u\in U$ and $\operatorname*{sat}(u)=0$ otherwise, and $\partial u^b/\partial K$ is given by
\begin{align*}
\frac{\partial u^b}{\partial K_t} = - \mathds{1}_t \otimes w^\top,
\end{align*}
for each $t$, where $\otimes$ denotes the Kronecker product and $\mathds{1}_t$ is the vector containing a one in its $t$-th row and $0$ elsewhere. 
In \cref{sec:results} we use Pytorch \cite{paszke2019pytorch} to efficiently evaluate the expression in \cref{eqn:FullGradient}.

Given the gradient formula in (24), we solve the unconstrained reformulation of (23) using stochastic gradient descent with the Adam algorithm \cite{kingma2014adam}. Specifically, at each iteration we randomly sample elements of $w$ and $\theta$, evaluate \cref{eqn:FullGradient} and implement a gradient step update. The complete training procedure is summarized in \cref{alg:backprop}, in which $m_k$ and $v_k$ denote the first and second moment vectors, initialized as $m_0=0$ and $v_0=0$, $\gamma>0$ is the learning rate, $\beta_1,\beta_2\in[0,1)$ are the exponential decay rates of the moment estimates, and $\epsilon>0$ is a small constant added to avoid division by zero.

\begin{algorithm} 
\caption{Gradient-based policy optimization algorithm to solve \cref{eqn:InLayerOptimization}.} \label{alg:backprop} 
\begin{algorithmic}[1] 
\Require Initial gain matrix $K^0$, learning rate $\gamma>0$, decay rates $\beta_1,\beta_2\in[0,1)$, numerical constant $\epsilon>0$, number of iterations $N>0$. 
\Init $m_0 \gets 0$, $v_0 \gets 0$. 
\For{$k=0,1,\dots,N-1$} 
\State Sample $w[t] \sim \operatorname*{Uniform}[w_\text{min},w_\text{max}]$ for $t=1,\dots,t_p$. 
\State Sample $\theta\sim \operatorname*{Uniform}[\theta_\text{min},\theta_\text{max}]$. 
\State Compute the stochastic gradient $g_k \gets \nabla J_\text{in}(K^k)$ using \cref{eqn:FullGradient}. 
\State $m_{k+1} \gets \beta_1 m_k + (1-\beta_1)g_k$. 
\State $v_{k+1} \gets \beta_2 v_k + (1-\beta_2)g_k^2$. \algorithmiccomment{Square applied elementwise.} 
\State $\hat m_{k+1} \gets m_{k+1}/(1-\beta_1^{k+1})$. 
\State $\hat v_{k+1} \gets v_{k+1}/(1-\beta_2^{k+1})$. 
\State $K^{k+1} \gets K^k - \gamma \frac{\hat m_{k+1}}{\sqrt{\hat v_{k+1}}+\epsilon}$. 
\EndFor 
\MyReturn $K^N$. 
\end{algorithmic} 
\end{algorithm}

\section{Experiment Design}
\label{sec:experiment_design}
This section is divided into three parts. \cref{sec:experiment_design:calibration} calibrates the control-oriented
thermal model and the affine sensor map to the physical LPBF system in the simulation environment, using data from a single uncontrolled layer. \cref{sec:experiment_design:experimental} describes the setup of hardware-compatible controller used in the physical study and its
offline simulation-based tuning. Finally, \cref{sec:experiment_design:plan} specifies the printed specimens and post-process
measurements used for the experimental comparison.


\subsection{Simulation-to-Reality Parameter Calibration}
\label{sec:experiment_design:calibration}

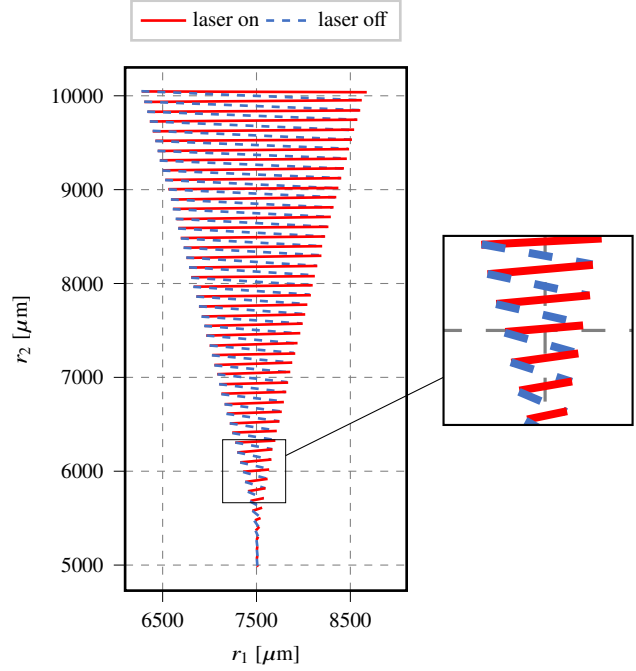
\begin{figure}
    \centering
\begin{tikzpicture}[spy using outlines={rectangle, magnification=3, size=2.5cm, connect spies}]

\definecolor{gray}{RGB}{128,128,128}
\definecolor{lightgray204}{RGB}{204,204,204}
\definecolor{steelblue68114196}{RGB}{68,114,196}

\begin{axis}[
scaled y ticks = false,
legend cell align={left},
legend columns=2,
legend style={
  fill opacity=0.8,
  draw opacity=1,
  text opacity=1,
  at={(0.5,1.13)},
  anchor=north,
  draw=lightgray204
},
tick align=outside,
tick pos=left,
x grid style={gray},
xlabel={$r_1$ [$\mu$m]},
xmin=6100, xmax=9100,
xtick style={color=black},
xtick={6500,7500,8500},
xticklabels={6500, 7500, 8500},
y grid style={gray},
ylabel={$r_2$ [$\mu$m]},
ymin=4727.82, ymax=10302.18,
ytick style={color=black},
ytick={5000,6000,7000,8000,9000,10000},
yticklabels={
    5000,
    6000,
    7000,
    8000,
    9000,
    10000
},
height=8.5cm, width=8.5cm,
line width=1,
grid style={darkgray176,dashed},
ymajorgrids, xmajorgrids,
label style = {font=\allfigurefontsize},
legend style = {font=\allfigurefontsize},
ticklabel style = {font=\footnotesize},
axis equal image
]
\addplot [red]
table {%
8674.8 10038
6277.8 10046.4
};
\addlegendentry{laser on}
\addplot [steelblue68114196, dashed]
table {%
6273.6 10048.8
8610 9953.4
};
\addlegendentry{laser off}
\addplot [red, forget plot]
table {%
8622 9952.8
6304.8 9934.2
};
\addplot [steelblue68114196, dashed, forget plot]
table {%
6303.6 9937.2
8592.6 9847.2
};
\addplot [red, forget plot]
table {%
8605.2 9844.2
6339.6 9829.2
};
\addplot [steelblue68114196, dashed, forget plot]
table {%
6334.8 9831
8566.2 9745.2
};
\addplot [red, forget plot]
table {%
8574.6 9745.2
6363 9726
};
\addplot [steelblue68114196, dashed, forget plot]
table {%
6364.2 9727.2
8538 9633.6
};
\addplot [red, forget plot]
table {%
8540.4 9639
6394.2 9621
};
\addplot [steelblue68114196, dashed, forget plot]
table {%
6393.6 9626.4
8506.8 9532.8
};
\addplot [red, forget plot]
table {%
8517 9532.8
6422.4 9520.2
};
\addplot [steelblue68114196, dashed, forget plot]
table {%
6421.8 9518.4
8472 9436.2
};
\addplot [red, forget plot]
table {%
8486.4 9432
6448.2 9412.2
};
\addplot [steelblue68114196, dashed, forget plot]
table {%
6445.2 9412.8
8456.4 9331.8
};
\addplot [red, forget plot]
table {%
8462.4 9329.4
6471 9312.6
};
\addplot [steelblue68114196, dashed, forget plot]
table {%
6468 9310.2
8423.4 9224.4
};
\addplot [red, forget plot]
table {%
8431.8 9226.8
6502.8 9205.8
};
\addplot [steelblue68114196, dashed, forget plot]
table {%
6498 9203.4
8397 9123
};
\addplot [red, forget plot]
table {%
8403 9122.4
6528.6 9103.8
};
\addplot [steelblue68114196, dashed, forget plot]
table {%
6526.2 9102.6
8366.4 9013.8
};
\addplot [red, forget plot]
table {%
8372.4 9018
6559.8 9004.8
};
\addplot [steelblue68114196, dashed, forget plot]
table {%
6566.4 9001.8
8336.4 8918.4
};
\addplot [red, forget plot]
table {%
8350.2 8919.6
6590.4 8894.4
};
\addplot [steelblue68114196, dashed, forget plot]
table {%
6587.4 8895.6
8310.6 8809.2
};
\addplot [red, forget plot]
table {%
8321.4 8812.2
6612 8794.8
};
\addplot [steelblue68114196, dashed, forget plot]
table {%
6607.2 8792.4
8285.4 8709.6
};
\addplot [red, forget plot]
table {%
8292 8709.6
6647.4 8686.2
};
\addplot [steelblue68114196, dashed, forget plot]
table {%
6637.8 8689.2
8253 8604
};
\addplot [red, forget plot]
table {%
8266.2 8604
6667.8 8590.8
};
\addplot [steelblue68114196, dashed, forget plot]
table {%
6664.2 8586
8222.4 8501.4
};
\addplot [red, forget plot]
table {%
8231.4 8500.8
6691.8 8484.6
};
\addplot [steelblue68114196, dashed, forget plot]
table {%
6694.8 8480.4
8191.8 8398.2
};
\addplot [red, forget plot]
table {%
8199.6 8397.6
6725.4 8381.4
};
\addplot [steelblue68114196, dashed, forget plot]
table {%
6722.4 8381.4
8178.6 8301
};
\addplot [red, forget plot]
table {%
8191.8 8295.6
6753.6 8272.8
};
\addplot [steelblue68114196, dashed, forget plot]
table {%
6751.2 8277
8141.4 8188.8
};
\addplot [red, forget plot]
table {%
8148 8187
6781.8 8169.6
};
\addplot [steelblue68114196, dashed, forget plot]
table {%
6780.6 8172.6
8110.8 8088
};
\addplot [red, forget plot]
table {%
8121.6 8078.4
6804 8066.4
};
\addplot [steelblue68114196, dashed, forget plot]
table {%
6805.2 8061.6
8088.6 7981.8
};
\addplot [red, forget plot]
table {%
8098.8 7981.2
6831.6 7965.6
};
\addplot [steelblue68114196, dashed, forget plot]
table {%
6825.6 7965
8067.6 7881.6
};
\addplot [red, forget plot]
table {%
8079.6 7875
6859.8 7861.2
};
\addplot [steelblue68114196, dashed, forget plot]
table {%
6858 7861.2
8034.6 7781.4
};
\addplot [red, forget plot]
table {%
8041.8 7774.8
6886.8 7754.4
};
\addplot [steelblue68114196, dashed, forget plot]
table {%
6886.8 7757.4
8005.8 7675.2
};
\addplot [red, forget plot]
table {%
8019 7670.4
6918 7648.2
};
\addplot [steelblue68114196, dashed, forget plot]
table {%
6913.2 7647.6
7980 7573.8
};
\addplot [red, forget plot]
table {%
7992.6 7572.6
6948 7548
};
\addplot [steelblue68114196, dashed, forget plot]
table {%
6942 7548.6
7953 7467.6
};
\addplot [red, forget plot]
table {%
7966.8 7470.6
6964.2 7444.8
};
\addplot [steelblue68114196, dashed, forget plot]
table {%
6967.2 7446
7931.4 7360.8
};
\addplot [red, forget plot]
table {%
7937.4 7365
7000.2 7337.4
};
\addplot [steelblue68114196, dashed, forget plot]
table {%
7002.6 7339.8
7898.4 7260
};
\addplot [red, forget plot]
table {%
7912.8 7254
7029 7237.2
};
\addplot [steelblue68114196, dashed, forget plot]
table {%
7024.2 7235.4
7875 7157.4
};
\addplot [red, forget plot]
table {%
7881 7157.4
7050.6 7131.6
};
\addplot [steelblue68114196, dashed, forget plot]
table {%
7051.8 7128
7852.2 7053.6
};
\addplot [red, forget plot]
table {%
7860 7057.8
7081.2 7033.8
};
\addplot [steelblue68114196, dashed, forget plot]
table {%
7084.2 7032.6
7825.8 6949.2
};
\addplot [red, forget plot]
table {%
7837.2 6944.4
7105.8 6926.4
};
\addplot [steelblue68114196, dashed, forget plot]
table {%
7101 6927.6
7809.6 6843
};
\addplot [red, forget plot]
table {%
7817.4 6844.8
7125.6 6825.6
};
\addplot [steelblue68114196, dashed, forget plot]
table {%
7126.2 6820.8
7783.8 6743.4
};
\addplot [red, forget plot]
table {%
7791.6 6736.8
7161 6714
};
\addplot [steelblue68114196, dashed, forget plot]
table {%
7159.2 6714
7764 6636.6
};
\addplot [red, forget plot]
table {%
7769.4 6633.6
7186.8 6613.8
};
\addplot [steelblue68114196, dashed, forget plot]
table {%
7179 6612
7741.8 6532.2
};
\addplot [red, forget plot]
table {%
7744.8 6531
7215.6 6508.8
};
\addplot [steelblue68114196, dashed, forget plot]
table {%
7215 6510.6
7714.2 6432.6
};
\addplot [red, forget plot]
table {%
7716 6429
7241.4 6411
};
\addplot [steelblue68114196, dashed, forget plot]
table {%
7243.2 6406.2
7696.8 6332.4
};
\addplot [red, forget plot]
table {%
7701 6326.4
7273.2 6304.2
};
\addplot [steelblue68114196, dashed, forget plot]
table {%
7275 6306
7668 6234.6
};
\addplot [red, forget plot]
table {%
7669.8 6232.8
7295.4 6199.8
};
\addplot [steelblue68114196, dashed, forget plot]
table {%
7295.4 6201
7651.2 6124.2
};
\addplot [red, forget plot]
table {%
7659.6 6124.8
7326.6 6094.2
};
\addplot [steelblue68114196, dashed, forget plot]
table {%
7326 6093.6
7633.8 6023.4
};
\addplot [red, forget plot]
table {%
7635 6018
7358.4 5994.6
};
\addplot [steelblue68114196, dashed, forget plot]
table {%
7356.6 5992.2
7609.2 5920.8
};
\addplot [red, forget plot]
table {%
7618.2 5918.4
7380 5885.4
};
\addplot [steelblue68114196, dashed, forget plot]
table {%
7378.2 5887.2
7594.8 5817.6
};
\addplot [red, forget plot]
table {%
7596 5818.8
7408.8 5786.4
};
\addplot [steelblue68114196, dashed, forget plot]
table {%
7407 5782.2
7575.6 5710.2
};
\addplot [red, forget plot]
table {%
7578 5713.2
7436.4 5683.8
};
\addplot [steelblue68114196, dashed, forget plot]
table {%
7430.4 5679.6
7556.4 5610.6
};
\addplot [red, forget plot]
table {%
7560 5611.2
7456.2 5577.6
};
\addplot [steelblue68114196, dashed, forget plot]
table {%
7456.8 5578.8
7540.8 5509.8
};
\addplot [red, forget plot]
table {%
7543.2 5501.4
7483.8 5476.8
};
\addplot [steelblue68114196, dashed, forget plot]
table {%
7476 5473.2
7522.8 5407.2
};
\addplot [red, forget plot]
table {%
7530 5407.2
7491.6 5367
};
\addplot [steelblue68114196, dashed, forget plot]
table {%
7489.8 5366.4
7517.4 5304.6
};
\addplot [red, forget plot]
table {%
7518 5301
7500 5267.4
};
\addplot [steelblue68114196, dashed, forget plot]
table {%
7500 5266.2
7509 5196.6
};
\addplot [red, forget plot]
table {%
7509.6 5199
7503.6 5161.2
};
\addplot [steelblue68114196, dashed, forget plot]
table {%
7504.2 5162.4
7511.4 5093.4
};
\addplot [red, forget plot]
table {%
7509.6 5096.4
7502.4 5065.8
};
\addplot [steelblue68114196, dashed, forget plot]
table {%
7503.6 5067
7509.6 4991.4
};
\addplot [red, forget plot]
table {%
7508.4 4991.4
7515 4981.2
};
\coordinate (spypoint) at (axis cs:7475,6000);
\coordinate (spyviewer) at (axis cs:10500,7500);
\end{axis}

\spy [black, line width = 0.5pt] on (spypoint) in node[anchor=center] at (spyviewer);

\end{tikzpicture}
    \caption{Wedge path for the laser. The laser scans from right to left, with hatching proceeding from top to bottom.}
    \label{fig:WedgePath}
\end{figure}

For the physical study, we applied the dynamical model introduced in \cref{sec:ControlOrientedMultiLayerModel} to the wedge path of \cite{kavas2024situ}, shown in \cref{fig:WedgePath}. This path exhibits significant in-layer overheating. We calibrated five parameters: the absorbed-power fraction $\alpha$, the laser-beam radius $a$, the porosity $\epsilon$, the thermal conductivity $\kappa_\text{pd}$ between powder and solidified metal, and the thermal conductivity $\kappa_\text{p}$ of the powder. Their search intervals are listed in \cref{tab:grid_search}.

The calibration data, denoted by $y_\text{real}$, consist of the pyrometer signal in millivolts obtained from a single uncontrolled layer of the wedge print with a constant laser power of $150\,\si{\watt}$. For a candidate parameter vector $\theta=[\alpha,a,\epsilon,\kappa_\text{pd},\kappa_\text{p}]^\top$, let $y_\text{K}(\theta)$ denote the corresponding simulated output in Kelvin. Since the simulated temperature and the measured signal have different units, we jointly calibrated the control-oriented parameters and an affine sensor map:
\begin{align*}
\left(\theta^{\mathrm{n}},A^{\mathrm{n}},b^{\mathrm{n}}\right)
&\in
\operatorname*{arg\,min}_{\substack{\theta\in\Theta_\text{grid}\\ A,b}}
\left\|y_\text{real}-\left(Ay_\text{K}(\theta)+b\mathbf{1}\right)\right\|_2^2.
\end{align*}
Here, $\Theta_\text{grid}$ is the Cartesian product of the parameter samples in \cref{tab:grid_search}. The joint problem was evaluated by enumerating $\theta\in\Theta_\text{grid}$, solving the linear least-squares problem for $A$ and $b$ at each grid point, and retaining the tuple with the smallest residual. The resulting $A^{\mathrm{n}}$ and $b^{\mathrm{n}}$ were fixed thereafter, and the sensor-domain model output is defined as
\begin{equation*}
y(\theta)=A^{\mathrm{n}}y_\text{K}(\theta)+b^{\mathrm{n}}\mathbf{1}.
\end{equation*}
The nominal values $\theta^{\mathrm{n}}$ are reported in \cref{Tab:ProcessParametersReal,Tab:ProcessParametersRealCalibration} of \ref{sec:ModelParametersForSimulationAndExperiments}.

\begin{table}[H]
\small
\caption{Uncertain parameters and the intervals used in the grid search.}
\label{tab:grid_search}
\centering
\begin{tabular}{c c c c}
\toprule
\textbf{Parameter} & \textbf{Description} & \textbf{Interval} & \textbf{Number of samples} \\
\midrule
$\alpha$      & Fraction of laser power absorbed by the material          & $[0.5,1]$                    & 6  \\
$a$           & Cross-section radius of the laser beam                     & $[4,9]\cdot 10^{-4}$\,\si{\meter} & 6  \\
$\epsilon$    & Porosity of the metal                                      & $[0.4,0.6]$                  & 5  \\
$\kappa_\text{pd}$ & Thermal conductivity between powder and solidified metal & $[0.67,1.5]$\,\si{\watt\per\meter\per\kelvin} & 6  \\
$\kappa_\text{p}$  & Thermal conductivity of powder                           & $[0.5,5]$\,\si{\watt\per\meter\per\kelvin} & 10 \\
\bottomrule
\end{tabular}
\end{table}

\begin{figure}
    \centering
    \input{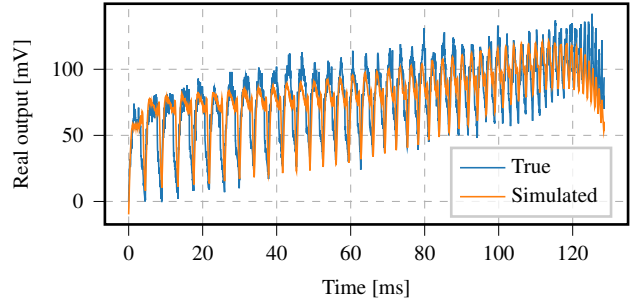}\\
    \caption{Calibrated simulation output $y(\theta^{\mathrm{n}})$ (orange) and measured output $y_\text{real}$ (blue) for the wedge path in \cref{fig:WedgePath} with a constant input of $150\,\si{\watt}$.}
    \label{fig:SimRealSimilarity}
\end{figure}

The measured output and the calibrated simulation output are compared in \cref{fig:SimRealSimilarity}. Note that \emph{the tuning was performed in simulation, requiring only the data of a single-layer laser exposure on the real machine for calibration, thus significantly reducing costs.} These calibrated quantities were used for constructing the simulation environment for the hardware-compatible controller tuning described in \cref{sec:experiment_design:experimental}.

\subsection{Physical Experimental Setup and Controller Tuning}
\label{sec:experiment_design:experimental}

\begin{figure}[b]
\centering
\scalebox{0.9}{
    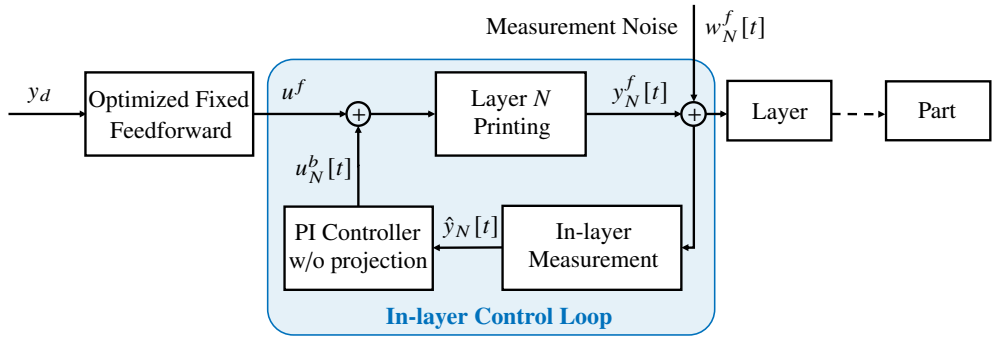
}
\caption{Simplified control architecture used for the physical LPBF experiments.}
\label{fig:SLMSimplifiedControlArchitecture}
\end{figure}

The calibrated model from \cref{sec:experiment_design:calibration} provides the simulation environment
for offline tuning of the controller used in the physical wedge study.
Relative to the complete multi-scale control architecture introduced in \cref{sec:methods_control}, the available LPBF hardware requires two adaptations, summarized in \cref{fig:SLMSimplifiedControlArchitecture}:
\begin{itemize}
    \item The setup is not capable of run-time adjustments of layer-to-layer parameters together with the PI controller, which is a common limitation with existing machines. Therefore, we designed the layer-to-layer control input as a fixed feedforward signal for the PI controller.
    \item The input projection for the in-layer controller is not utilized in this study to avoid nonlinearities caused by input saturation and the absence of anti-windup implementations on the physical setup. Our controller tuning keeps the laser input within the desired process window with minimal violations during transients (beginning/end of vector). We experimentally verified that clipping the inputs within a narrow process window band results in excessive lack of fusion.
\end{itemize}

The layer-to-layer term is therefore $u_N^f[t]=u^f$ for every printed layer $N$ and $t=0,\ldots,t_p-1$. The in-layer loop is implemented as a proportional-integral (PI) controller,
\begin{align*}
u_N^b[t]
    &=K_p e_N[t]+K_i\Delta t\sum_{i=0}^{t}e_N[i],\\
u_N[t]
    &=u^f+u_N^b[t],
\end{align*}
where the integral term is accumulated only while the laser was on. With these adaptations, the implemented controller is a PI controller with a fixed feedforward term, a structure related to those in \cite{craeghs2010feedback,shkoruta2022real,kavas2024situ}. Here, however, \emph{the controller is only tuned in simulation using a model calibrated from a single-layer exposure on the physical machine.}

We retained the simulation-based training principle introduced in \cref{sec:InLayerControlLoopDesign}, but optimized $K=(K_p,K_i)$ and $u^f$ on the calibrated wedge model. Because the physical implementation does not use $\Pi_U$, the input limits are incorporated through penalty terms. The training objective is
\begin{equation}
J_{\text{in}}(K,u^f)
=\frac{1}{t_p}\mathbb{E}_{w,\theta'}\!\left[
\sum_{j=1}^{t_p}\norm{y_1[j]-y_d}^2
+\lambda P_\text{cst}(u_1)
+\eta P_\text{osc}(u_1)
\right],
\label{eq:cost_nominal}
\end{equation}
where $P_\text{cst}$ penalizes input-constraint violations and $P_\text{osc}$ penalizes rapid input variations:
\begin{align*}
P_\text{cst}(u_1)
    &=\sum_{j=0}^{t_p-1}\left(
        \bigl[u_\text{min}-u_1[j]\bigr]_+
        +\bigl[u_1[j]-u_\text{max}\bigr]_+
      \right),\\
P_\text{osc}(u_1)
    &=\sum_{j=1}^{t_p-2}\left(
        u_1[j-1]-2u_1[j]+u_1[j+1]
      \right).
\end{align*}
Here, $[x]_+=\max\{0,x\}$, and $\lambda$ and $\eta$ are penalty weights. The first penalty $P_\text{cst}$ encourages the laser power to remain in $[u_\text{min},u_\text{max}]$, reducing the risks of lack of fusion below $u_\text{min}$ and keyholing above $u_\text{max}$. The second penalty $P_\text{osc}$ discourages aggressive changes between successive time steps in the control input.

The complete tuning problem is
\begin{equation}
\begin{aligned}
\operatorname*{minimize}_{K,u^f}\quad
    &J_{\text{in}}(K,u^f)
    =\frac{1}{t_p}\mathbb{E}_{w,\theta'}\!\left[
        \sum_{j=1}^{t_p}\norm{y_1[j]-y_d}^2
        +\lambda P_\text{cst}(u_1)
        +\eta P_\text{osc}(u_1)
      \right]\\
\text{subject to}\quad
    &y_1=Y_1^u(\theta')u_1+Y_1^x(\theta')X_1[0]
        +Y_1^d(\theta')\bar d_1(\theta'),\\
    &\hat y_1[t]=y_1[t]+w[t],\\
    &w[t]\sim\text{Uniform}[w_\text{min},w_\text{max}],\\
    &\theta'\sim\text{Uniform}[\theta_\text{min},\theta_\text{max}],\\
    &u_1[t]=u^f+u_1^b[t],\qquad t=0,1,\ldots,t_p-1,\\
    &u_1^b[t]=K_p e_1[t]+K_i\Delta t\sum_{i=0}^{t}e_1[i],
        \qquad t=0,1,\ldots,t_p-1,\\
    &e_1[i]=y_1[i]-\hat y_1[i],\qquad i=0,1,\ldots,t,\\
    &X_1[0]=\bar X_1[0].
\end{aligned}
\label{eq:wedge_design_problem}
\end{equation}

We optimized $K_p$, $K_i$, and $u^f$ by gradient descent. At each iteration, we sample
$\theta'=\operatorname{diag}(1+e_\theta)\theta^{\mathrm{n}}$, where the entries of $e_\theta$ are independently drawn from $\text{Uniform}[-0.05,0.05]$. Measurement noise is sampled independently at each time step from the interval $[-5,5]$. We ran the Adam optimizer for 200 iterations with $\beta_1=0$ and $\beta_2=0.8$. The learning rates for $K_p$, $K_i$, and $u^f$ are $4\cdot10^{-3}$, $5\cdot10^{-4}$, and $2$, respectively. Using the fixed affine map from \cref{sec:experiment_design:calibration}, the desired output in Kelvin corresponding to the $\SI{80}{\milli\volt}$ reference is $y_d=(\SI{80}{\milli\volt}-b^{\mathrm{n}})/A^{\mathrm{n}}$.

\subsection{Physical Experiment Plan}
\label{sec:experiment_design:plan}

The physical study comprised the following specimen groups:
\begin{itemize}
    \item \textit{Proposed controllers with rate penalization:}
    $\eta \neq 0$. These specimens are included in the main time-series comparison in \cref{fig:RealComparison}, the tracking error and constraint violation comparison in \cref{tab:more_experiments}, and the spatial, microstructural, geometric, and high-frequency comparisons in \cref{fig:RealComparisonScatter,fig:3d_images_isometric,fig:results:experimental:wedge_experiment_vectors}.

    \item \textit{Proposed controllers without rate penalization:}
    $\lambda=\{100, 125, 150\}$ and $\eta=0$. These specimens are used to investigate the effect of the oscillatory control
    signal in \cref{fig:RealComparisonScatter,fig:3d_images_isometric,fig:results:experimental:wedge_experiment_vectors,fig:porosity}.

    \item \textit{Bayesian optimization (BO) baseline:}
    the PI controller from \cite{kavas2024situ}, tuned without a
    static feedforward term by minimizing a combination of reference
    tracking error and successive input changes.

    \item \textit{Uncontrolled baseline:}
    a constant laser-power input of $150\,\mathrm{W}$.
\end{itemize}


The parts were printed on large cross-sectioned support structures, with each part comprising 150 layers. After removal from the build platform, the parts were scanned in 3D and imaged from multiple angles using a Keyence VHX-7000 optical microscope. Specimen preparation and microscopic evaluation followed the procedure in \cref{section:micro_evaluation}. Finally, the polished mounts were inspected under coaxial and ring illumination to characterize the microstructure.

\section{Materials}
\label{sec:materials}
All physical experiments were conducted on an Aconity3D Midi+ laser powder bed fusion (LPBF) system (Aconity3D GmbH, Herzogenrath, Germany) \cite{aconity3d}.
The setup employed a continuous-wave, Gaussian-mode fiber laser (nLIGHT Alta, Vancouver, WA, USA) with a wavelength of 1080 nm and a maximum power output of 500 W.
The laser was focused to a beam diameter of $\SI{80}{\micro\meter}$, and the layer thickness was set to $\SI{30}{\micro\meter}$.
For this configuration, the nominal process parameters ensuring full density were a laser power of 150 W and a scan speed of 800 mm/s.
Gas-atomized stainless steel 316L (1.4404) powder with a particle size distribution of $15$–$\SI{45}{\micro\meter}$ (CT POWDERRANGE 316LF, Carpenter Additive, Cheshire, UK) was used as the feedstock material.

\begin{figure}
     \centering
     \footnotesize
        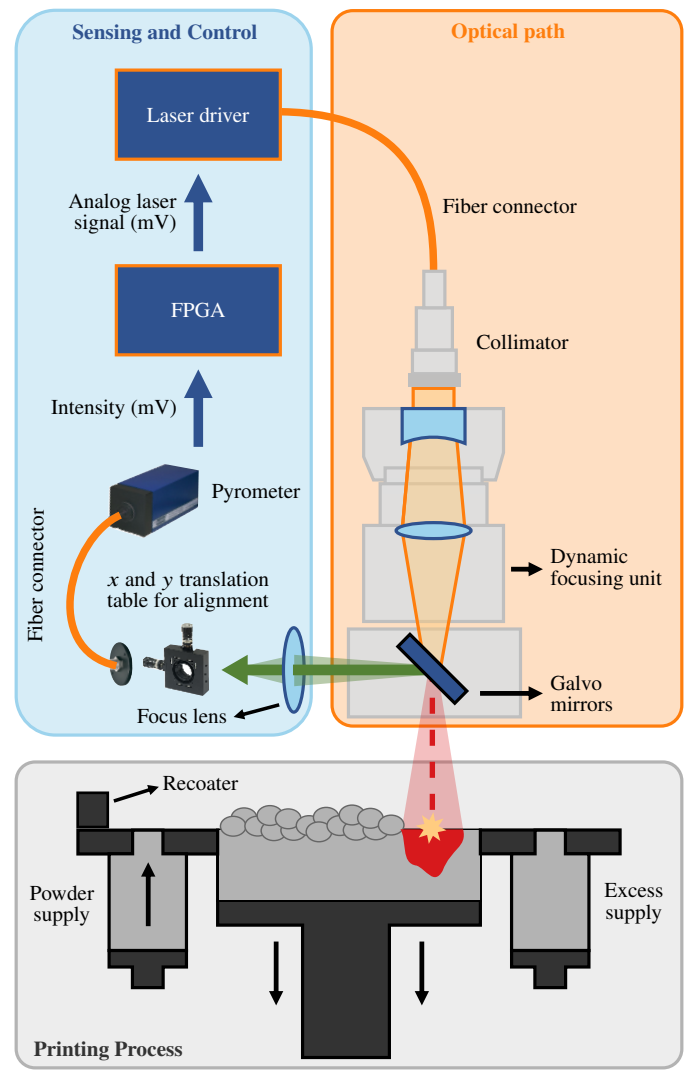
     \caption{The utilized hardware setup.}
     \label{fig:SLM_hardware_setup}
\end{figure}

The hardware architecture is summarized in \cref{fig:SLM_hardware_setup}.
The in-layer PI controller is implemented using the AconityCONTROL system.
The laser beam, transmitted via fiber optics, is collimated, expanded, and directed by a dichroic mirror to the galvanometer scanner.
Melt-pool radiation returns through the same path to a Kleiber KG740 pyrometer~\cite{kleiberinfrared} (1500–$\SI{1700}{\nano\meter}$), sampled at 100 kHz.
The signal is processed in an FPGA for PI control and laser power feedback, with a total latency of $\SI{20.279}{\micro\second}$, corresponding to a 50 kHz laser update rate (one update per two pyrometer samples).

\paragraph{Microscopic Evaluation}\label{section:micro_evaluation}
The specimens were manually detached from the build platform and cut near the center plane parallel to the build plate by a Struers Accutom-10 cutter.
They were then metallographically prepared by hot embedding in DuroLite bakelite using a Struers CitoPress, followed by sequential grinding with 320-grit sandpaper and polishing on Struers Largo, Dac, Nap, and Chem cloths using suspensions of $\SI{9}{\micro\meter}$, $\SI{3}{\micro\meter}$, $\SI{1}{\micro\meter}$, and $\SI{0.1}{\micro\meter}$ particles.
The polished samples were examined with a Keyence VHX-7000 digital microscope under coaxial and ring illumination.

\section{Results and Discussion}\label{sec:results}

This section reports the results in three stages. \cref{sec:results:numerical} evaluates the complete dual-loop
controller in a fully simulated setting under model uncertainty and
measurement noise. \cref{sec:results:wedge_tuning} presents the offline tuning results
for the hardware-compatible wedge controller introduced in \cref{sec:experiment_design:experimental}.
Finally, \cref{sec:results:physical} reports the corresponding physical
LPBF experiments.

\subsection{Numerical Evaluation of the Full Dual-Loop Controller}
\label{sec:results:numerical}

\subsubsection{Simulation Setup and Controller Training}
\label{sec:results:numerical_setup}

To evaluate the complete dual-loop architecture without the hardware limitations of the physical experiments, we conducted a fully simulated LPBF study using the multi-layer model.
This simulation employed a scan path and parameter set distinct from the hardware setup in \cref{sec:experiment_design:calibration}.

The fraction of absorbed laser power $\alpha$, the metal porosity $\epsilon$, and the thermal conductivity $\kappa_\text{pd}$ between powder and solidified metal are treated as uncertain parameters.
The parameter vector is defined as
\begin{equation*}
\theta=
\left[
\alpha'(1+e_\alpha),
\epsilon'(1+e_\epsilon),
\kappa'_\text{pd}(1+e_\kappa)
\right]^\top,
\end{equation*}
and its nominal value is $\theta'=[\alpha',\epsilon',\kappa_\text{pd}']^\top$. The exact numerical values are listed in \cref{Tab:ProcessParameters} of \ref{sec:ModelParametersForSimulationAndExperiments}.

\begin{figure}
\centering
\input{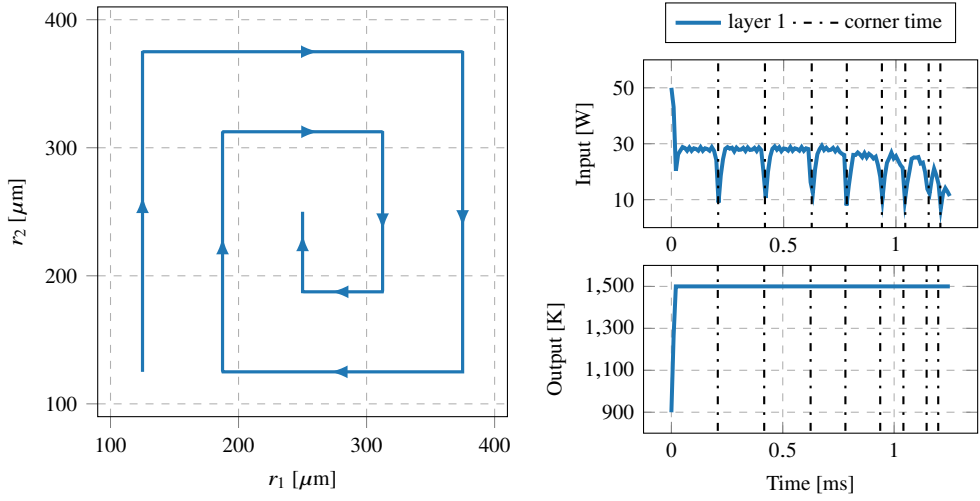}
\caption{Square-spiral laser path (left), associated optimal input (upper right), and realized output (lower right) for $y_d=1500~\text{K}$ and nominal parameters. The optimal laser power decreases sharply at each corner to stabilize the temperature.}
\label{fig:SquareSpiral}
\end{figure}

Both training and testing used the square-spiral scan path shown on the left of \cref{fig:SquareSpiral}. This path is challenging because it exhibits severe overheating and requires substantial input changes near its corners to stabilize the temperature, as illustrated on the right of \cref{fig:SquareSpiral}.

\paragraph{In-layer controller training}
\label{sec:TrainingSettingForInLayerControlLoop}
We trained the feedback gain $K$ by solving \cref{eqn:InLayerOptimization} using \cref{alg:backprop}. At each training iteration, a batch of 32 models is constructed by independently sampling $e_\alpha,e_\epsilon,e_\kappa\sim\text{Uniform}[-0.2,0.2]$. Starting from $K=0$, the controller is trained for 50 iterations using Adam with learning rate $\gamma=2\cdot10^{-3}$, $\beta_1=0$, and $\beta_2=0.8$ (hyperparameters selected by manual tuning). Training takes approximately 3 minutes on an NVIDIA GeForce RTX 3050 Ti GPU. At each time step, measurement noise $w\sim\text{Uniform}[\SI{-10}{\kelvin},\SI{10}{\kelvin}]$ is sampled to improve robustness. The desired temperature is $y_d=\SI{1500}{\kelvin}$.

\subsubsection{Closed-Loop Robustness Results}
\label{sec:results:numerical_results}

We evaluated the trained dual-loop controller by simulating the LPBF
process for six consecutive layers. The relative parameter perturbations
$e_\alpha$, $e_\epsilon$, and $e_\kappa$ were selected from
$\mathcal{S}_e =
\{-0.20,-0.15,-0.10,-0.05,0,0.05,0.10,0.15,0.20\}.$
All combinations
$(e_\alpha,e_\epsilon,e_\kappa)\in\mathcal{S}_e^3$
were evaluated, resulting in $9^3=729$ test models.
At each time step, the measurement noise is sampled independently from
$w \sim \operatorname{Uniform}[-10\si{K},10\si{K}]$.

To isolate the contributions of the two control loops, we compared our dual-loop controller with both a single-loop in-layer controller (where $u_N^f[l] \equiv u_1^f[l]$ for all $N$ and $l$, and the full input is chosen as $u_N[l] = u_1^f[l]+ u_{N}^b[l]$), and a single-loop layer-to-layer controller without any in-layer feedback (where $K=0$ and the full input is $u_N[l]=u_N^f[l]$).

\begin{figure}[ht!]
     \centering
     \input{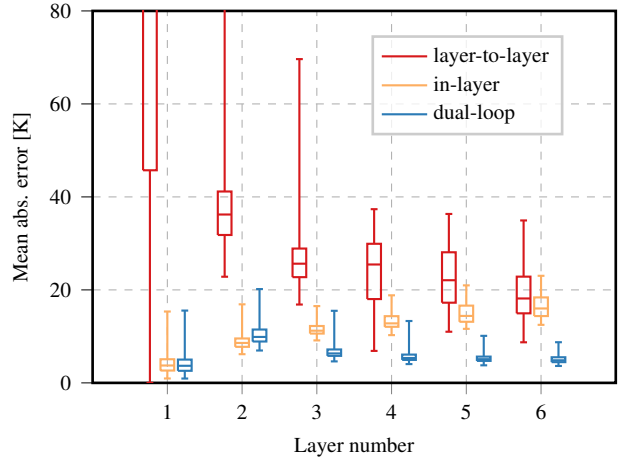}
     \vskip -3pt
     \caption{Box plot of the mean absolute error for each controller on all test models. 
     }
     \label{fig:MeanAE3controllers}
\end{figure}

Figure \ref{fig:MeanAE3controllers} shows the mean absolute error $\text{MeanAE}_k$ of different controllers across different layers, where
\begin{equation*}
    \text{MeanAE}_k = \frac{1}{t_p-2}\sum_{j=3}^{t_p} \norm{y_k[j] - y_d}.
\end{equation*}
\begin{figure}
     \centering
     \input{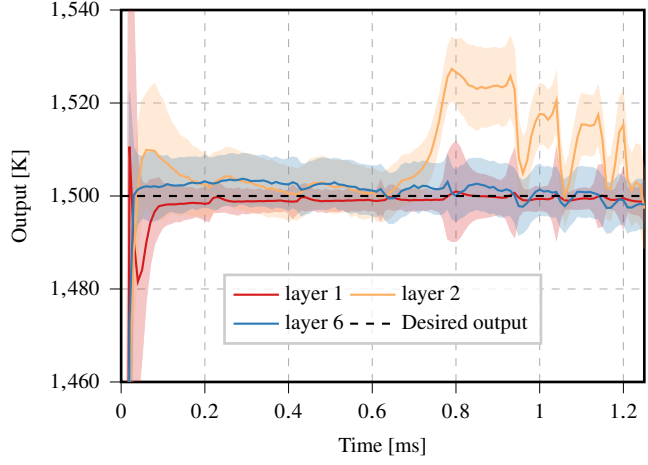}
     \vskip -3pt
     \caption{Mean (solid lines) and standard deviation (shaded areas) of all test models' outputs, obtained with the dual-loop controller.}
     \label{fig:Envelope}
\end{figure}

We chose $j=3$ as the starting time in $\text{MeanAE}_k$ because heating the powder layer from \SI{900}{\kelvin} to \SI{1500}{\kelvin} requires several time steps even when applying the maximal laser input. 
As a result, including the first few time steps in the statistics would raise the baseline error equally for all controllers, biasing the results.

\cref{fig:MeanAE3controllers} shows that the dual-loop controller outperforms the two single-loop controllers. 
Specifically, while the in-layer controller initially performs well, its mean absolute error shows an upward trend as more layers are printed, 
indicating that this control architecture alone struggles to maintain accuracy when heat starts accumulating across layers.

Figure \ref{fig:Envelope} reports the mean and standard deviation of all tested outputs obtained with the dual-loop controller across different layers. 
The output curve converges quickly to a neighborhood of the desired output in each layer, despite the presence of noise. 
In layer 2, a larger error can be observed in the second half of the scan due to the transition from single-layer to multi-layer dynamics, during which the model mismatch is most pronounced.
After this initial transition, the layer-to-layer correction adapts to the mismatch, reducing the error. 
Importantly, the maximum average deviation is only around $\SI{20}{\kelvin}$, corresponding to about $3\%$ of the relevant temperature range $\SI{900}{\kelvin}$-$\SI{1500}{\kelvin}$. 

In terms of computation, the in-layer linear feedback control has negligible computational cost as it only involves additions and multiplications. 
The layer-to-layer control loop takes up to \SI{0.02}{\second} to compute $u_N^f$ by using the Clarabel solver \cite{goulart2026clarabel}, making the scheme suitable for real-world deployment.


\subsection{Offline Tuning Results for the Hardware-Compatible Wedge Controller}
\label{sec:results:wedge_tuning}

\begin{figure}[hb!]
     \centering
\includegraphics{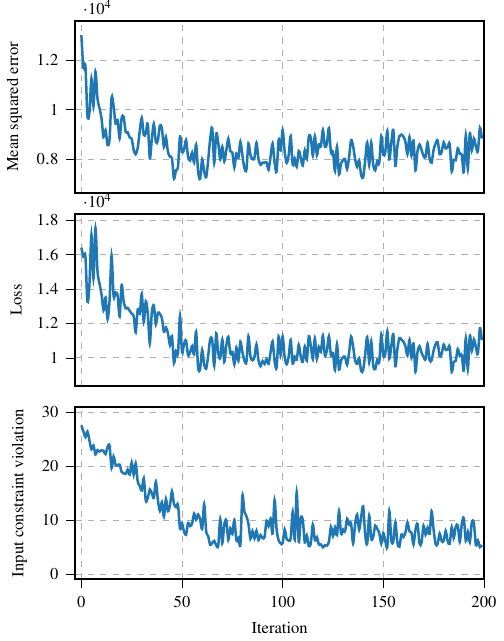}
     \caption{Mean squared error (top), loss (center), and constraint violation (bottom) across different iterations.}
     \label{fig:results:experimental:wedge_training}
\end{figure}

\begin{figure}[h]
     \centering
     \includegraphics{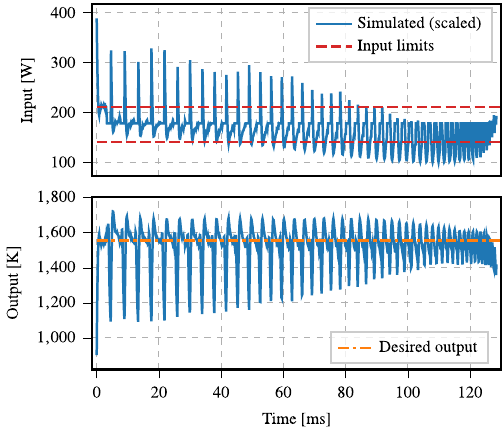}
     \caption{Simulation results for the wedge geometry. Input (above) and output (below) obtained with the trained controller parameters on the simulation 
     setup described in \cref{sec:experiment_design:experimental}.}
     \label{fig:results:experimental:wedge_simulation}
\end{figure}

We next report the offline tuning results for the hardware-compatible
controller introduced in \cref{sec:experiment_design:experimental}.
Unlike the complete dual-loop controller evaluated in
\cref{sec:results:numerical}, this controller uses a fixed feedforward
input and a PI feedback structure to accommodate the available LPBF
hardware.

We started by running \cref{alg:backprop} with the cost $J_\text{in}$ given in \cref{eq:cost_nominal} with $\lambda=30$
and $\eta=15$.
\cref{fig:results:experimental:wedge_training} shows how the mean squared error, the loss, and the input-constraint violation change across iterations
until the parameters converge to the optimal values $K_p=17.9$, $K_i=11303.2$, and $u^f=\SI{177.5}{\watt}$.
\cref{fig:results:experimental:wedge_simulation} shows the simulation input and output obtained with the tuned control parameters. 
For comparison purposes, we report here the parameters used in \cite{kavas2024situ}, obtained through BO: $K_p=8.45, K_i=90598.24$.

\subsection{Physical Experimental Results}
\label{sec:results:physical}

\begin{figure*}[b!]
     \centering
     \captionsetup{skip=0pt}
     \begin{subfigure}[c]{\textwidth}
          \centering
          \includegraphics[width=\columnwidth]{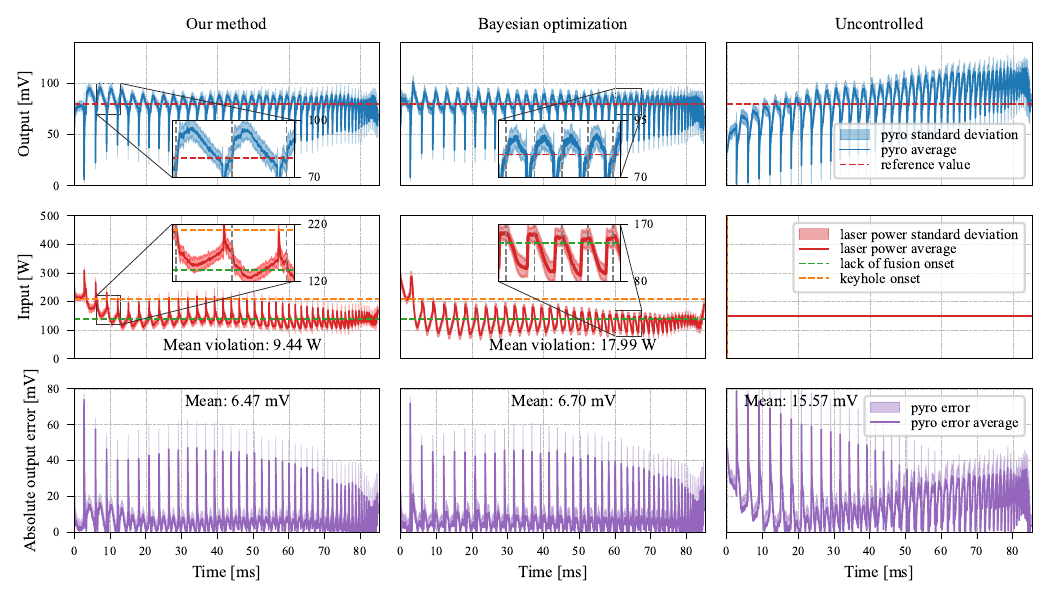}
     \end{subfigure}\\
     \caption{Time series comparison of pyrometer signals across wedge specimens using our controller (left column, with $\lambda=30, \eta=15$), offline-optimized BO (center column), and the uncontrolled setup (right column). 
     Shaded regions denote the standard deviation across layers. 
     The second row visualizes the power input profiles, 
     where the mean violation quantifies the average extent to which laser power exceeds the keyhole onset threshold or falls below the lack of fusion threshold.}%
     \label{fig:RealComparison}
\end{figure*}

We evaluated the hardware-compatible controller on the physical LPBF
system described in \cref{sec:materials}. The experimental results are shown in \cref{fig:RealComparison}. 
To improve the clarity, only the results for vectors where the laser is on are presented. 
The top row of subplots displays the pyrometer signal (in mV) across the printing window. 
The signal is averaged across layers $50$ to $80$. 
Observe how both our method (left) and BO in \cite{kavas2024situ} (center) effectively eliminate the severe overheating that occurs when a constant power profile is used (right).

The performance of our method closely matches that of BO, with a slight decrease in the average output error 
($6.47\,\text{mV}$ compared to $6.70\,\text{mV}$, as shown in the bottom row of subplots). 
Moreover, the power profile obtained with our method stays significantly closer to the desired operating range 
$140\,\text{W}$-$210\,\text{W}$ (the band between the dashed green and yellow lines in the second row of subplots), 
reducing the mean input violation error from $17.99\,\text{W}$ to $9.44\,\text{W}$. 
Although BO yields a smaller worst-case input violation, the average violation is the primary driving factor when evaluating the risks of both lack of fusion and keyholing.
By maintaining a significantly lower mean violation, our controller is expected to better prevent both defects, that is, preventing lack of fusion when the input drops below the process window (green line), and keyholing when it exceeds the process window (yellow line). 
Furthermore, by increasing the penalties on the input, the maximum violation can be reduced to a value smaller than BO, though at the cost of a slightly worsened tracking error.

In \cref{tab:more_experiments} we additionally tested different combinations of $\eta$ and $\lambda$, which led to similar results suggesting that the method is robust and does not require significant manual tuning of the penalty parameters.

\begin{table}
\footnotesize
\centering
\caption{Tracking error and constraint violation for all experiments.}
\begin{tabular}{ c c c c c c c c}
\toprule
\textbf{Type} & \multicolumn{2}{c}{\textbf{Weights}} & \multicolumn{3}{c}{\textbf{Cst. Violation}} & \multicolumn{2}{c}{\textbf{Error}} \\
\cmidrule(lr){2-3} \cmidrule(lr){4-6} \cmidrule(lr){7-8} & $\lambda$ & $\eta$ & \textit{Mean} & \textit{Std.} & \textit{Max} & \textit{Mean} & \textit{Std.} \\[0.25em]
Ours & 25 & 15   & 9.74\si{\watt} & 10.50\si{\watt} & 111.62\si{\watt} & 6.48\si{\milli\volt} & 5.11\si{\milli\volt} \\
     & 25 & 25   & 10.68\si{\watt}& 10.55\si{\watt} & 107.05\si{\watt} & \textbf{6.32\si{\milli\volt}} & 5.10\si{\milli\volt} \\
     & 30 & 15   & 9.44\si{\watt} & 9.99\si{\watt}  & 99.03\si{\watt}  & 6.47\si{\milli\volt} & 5.16\si{\milli\volt} \\
     & 50 & 50   & 7.39\si{\watt} & 8.05\si{\watt}  & 66.92\si{\watt}  & 7.11\si{\milli\volt} & 5.31\si{\milli\volt} \\
     & 50 & 75   & 7.18\si{\watt} & 7.97\si{\watt}  & 74.15\si{\watt}  & 7.45\si{\milli\volt} & 5.33\si{\milli\volt} \\
     & 75 & 50   & 6.37\si{\watt} & 7.38\si{\watt}  & \textbf{52.85\si{\watt}}  & 7.77\si{\milli\volt} & 5.39\si{\milli\volt} \\
     & 100 & 2.5 & \textbf{5.39\si{\watt}} & \textbf{6.80\si{\watt}}  & 57.76\si{\watt}  & 7.85\si{\milli\volt} & 5.46\si{\milli\volt} \\
BO   & NA & NA & 17.99\si{\watt} & 18.38\si{\watt}  & 76.62\si{\watt}  & 6.70\si{\milli\volt} & \textbf{5.03\si{\milli\volt}} \\
Uncontrolled & NA & NA & 0\si{\watt} & 0\si{\watt} & 0\si{\watt} & 15.57\si{\milli\volt} & 6.67\si{\milli\volt} \\
\bottomrule
\end{tabular}
\label{tab:more_experiments}
\end{table}



\begin{figure*}
     \centering
     \captionsetup{skip=0pt}
     \includegraphics[width=\textwidth]{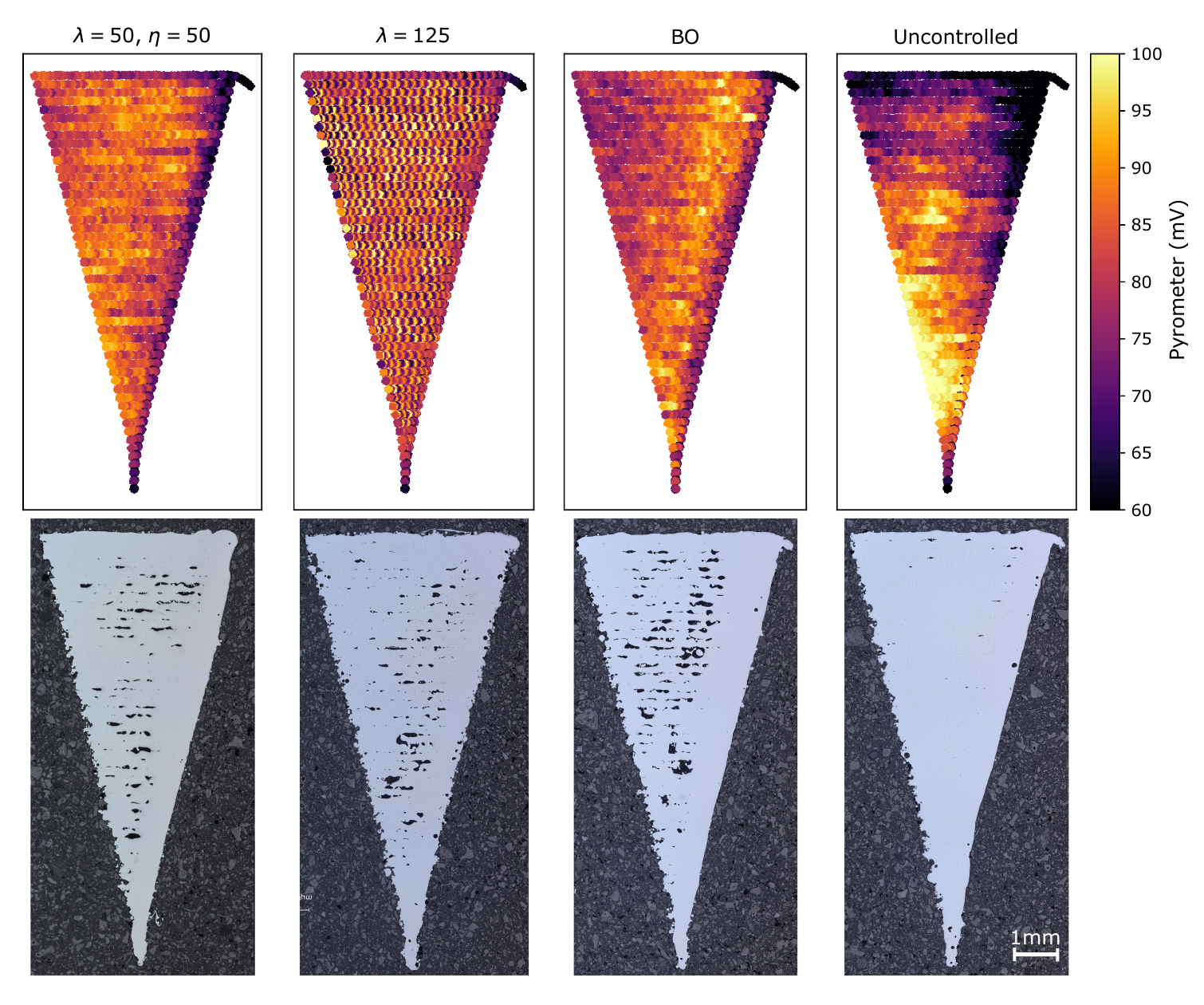}
     \caption{Scatter plot of the 4 cases. The first row shows the color map based on the pyrometer data, with all the controlled cases having 80mV as the reference value. In the second row, corresponding cut-ups are shown in an unetched condition for porosity evaluation.
     } \label{fig:RealComparisonScatter}
\end{figure*}

\cref{fig:RealComparisonScatter} compares in the first row the spatial distribution of pyrometer readings for (i) the BO, (ii) $\lambda = 125$, $\eta=0$, and (iii) $\lambda = 50,$ $\eta = 50$ cases. 
The scatter plots reveal how the measured surface temperatures evolve under different control strategies and parameter settings. 
Across the examined conditions, the color-coded deviations from the 80 mV reference value highlight regions of lower and higher than the expected range, enabling a direct visual assessment of thermal uniformity. 
In the second row, the corresponding microstructures from representative layers are shown.
The microstructures are shown in polished condition to observe the pores more clearly.


\begin{figure*}
     \centering
     \captionsetup{skip=0pt}
     \includegraphics[width=\textwidth]{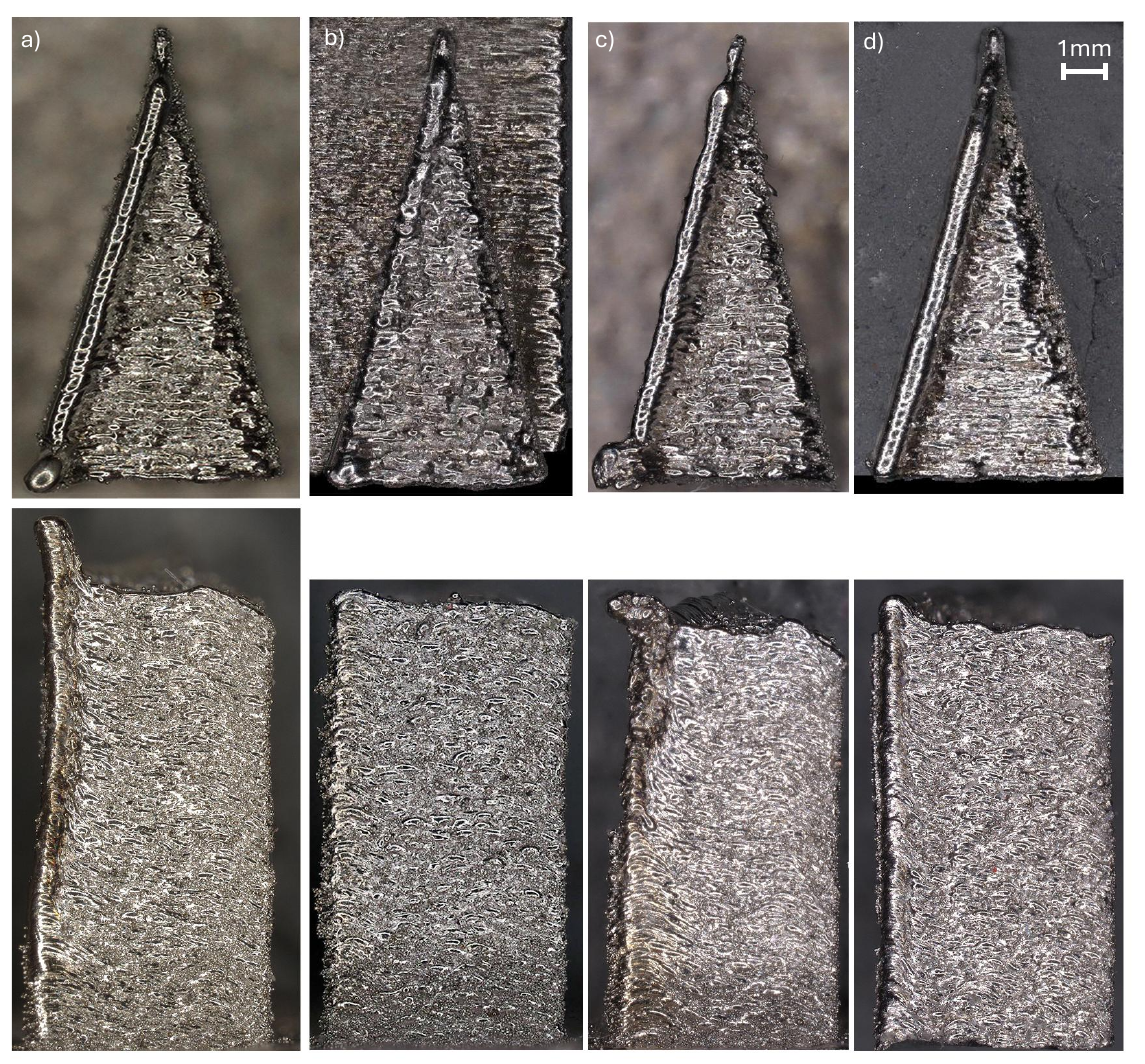}
     \caption{The upper surfaces are shown in the first row and the rear vertical surface of the wedges are in the second row. The columns show the a) $\lambda = 50~\ \eta = 50$, b) $\lambda = 125$ with no $\eta$, c) BO, and d) Uncontrolled cases.} 
     \label{fig:3d_images_isometric}
\end{figure*}

The printed parts are shown from the isometric angle in the first row of \cref{fig:3d_images_isometric} where the rear side of the wedge parts are shown in the second row, to observe the vector head swelling that is building up across the layers.


\begin{figure}
     \centering
     \includegraphics[width=0.9\columnwidth]{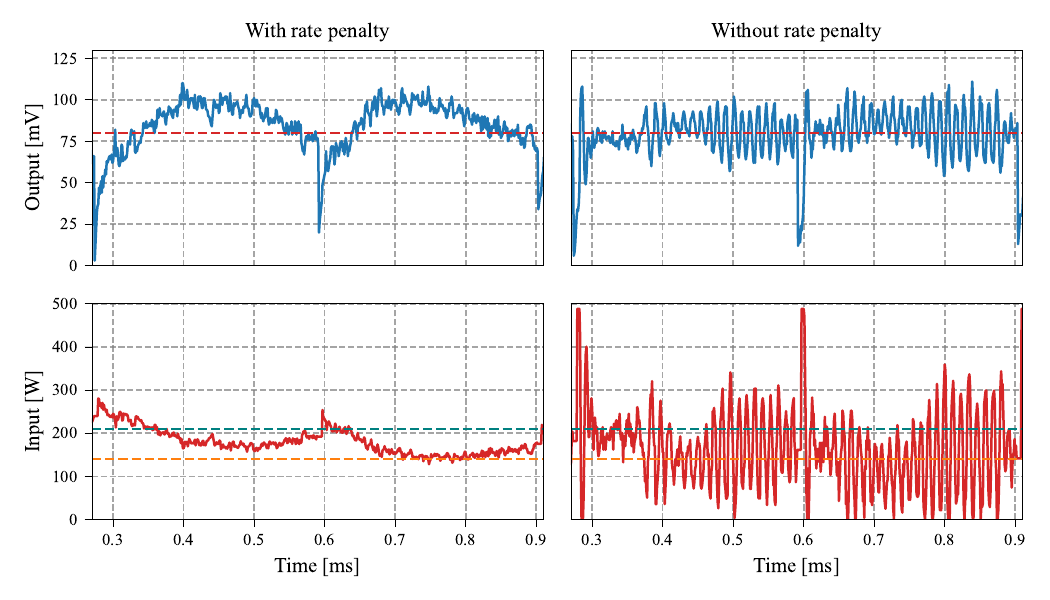}
     \caption{Experimental results of the input and output for the case with rate penalty $\lambda = 50, \eta = 50$ on the left, and the case without rate penalty $\lambda = 125, \eta=0$.}
     \label{fig:results:experimental:wedge_experiment_vectors}
\end{figure}


The results reveal that excessive energy input at vector initiation leads to pronounced swelling effects in both the BO and the $\lambda=50, \eta=50$ cases, particularly at vector beginnings as seen in \cref{fig:RealComparisonScatter}. 
This behavior is attributed to unidirectional hatch strategies, where overlapping melt pool initiation transients accumulate across layers, resulting in localized overheating and geometric distortion. 
These effects are clearly manifested in the upper surface profiles, where swelling is observed near vector start regions, as well as in the micrographs, which show enlarged grain structures. 
In contrast, although the vector beginning regions have a more uniform structure, $\lambda=125, \eta=0$ controlled part shows almost no swelling in the support surface profile, as observed in \cref{fig:3d_images_isometric}.

The first vector and its initiation region in the BO case exhibit exceptionally severe swelling, as evidenced by both micrographs and 3D isometric reconstructions, which is an effect that is not present to the same extent in the uncontrolled case. 
The proposed controller resolves this issue, eliminating the pronounced vector-start anomalies. A direct comparison of the length of the vector-initiation regions characterized by low porosity and large grains further demonstrates that the controller outperforms both the uncontrolled and BO strategies in stabilizing early vector behavior.

One notable observation when $\eta=0$ is that, despite the controller response inducing oscillatory behavior in both the laser power and the pyrometer signal (see \cref{fig:results:experimental:wedge_experiment_vectors}) with a 2 to 4 kHz frequency, a clear reduction in vector head swelling is observed (\cref{fig:3d_images_isometric}). 
When comparing the effects of the high frequency modulation, we see that while the case of no rate penalty in \cref{fig:results:experimental:wedge_experiment_vectors} shows larger input constraint violations, the resulting geometry in \cref{fig:3d_images_isometric} has favorable properties with reduced swelling, pointing to the need for further study.
Importantly, the effects of temporally modulated laser power have been investigated in prior work~\cite{caprio2024temporal,caprio2022understanding}, offering an interesting physical justification to this phenomenon.
The observed improvement may therefore be related to the influence of power modulation on the melt pool dynamics. 
In particular, temporal variations in the input power could affect local thermal gradients and fluid flow, potentially limiting the accumulation of melt at the vector head. 
The oscillatory inputs may also influence the stability of the vapor cavity and the associated recoil pressure, which in turn could suppress the development of surface instabilities. 
While this provides a possible explanation for the reduced swelling, the use of power input modulation within a closed-loop feedback control framework is a very promising direction for research in improving melt pool stability.

In the uncontrolled case, the tip region deviates from the intended geometry, protruding beyond the designed contour due to vector-initiation swelling. 
This deviation is effectively eliminated in the controlled builds, highlighting the controller's ability to more accurately reproduce the target geometry, which is an outcome of particular relevance for precision-critical additive manufacturing applications.

\begin{figure*}
     \centering
     \captionsetup{skip=0pt}
     \includegraphics[width=0.6\textwidth]{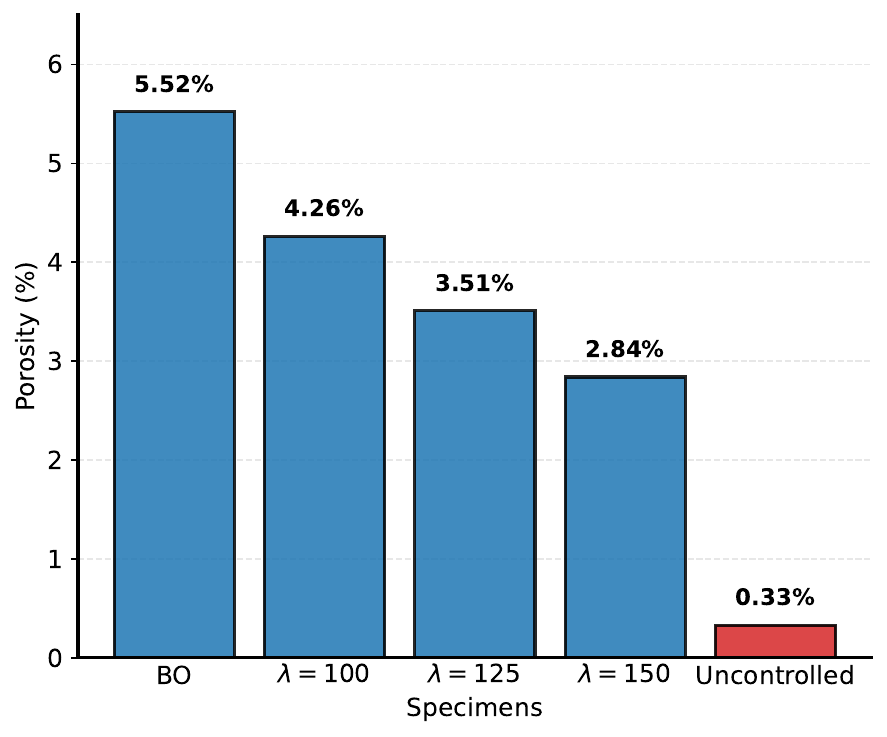}
     \caption{
     Comparison of porosity (\%) across wedges with different control strategies. The controlled cases reduce porosity from 5.52\% (BO) to 2.84\% at $\lambda = 150$. The uncontrolled specimen (0.33\%) is shown as a reference baseline.} \label{fig:porosity}
\end{figure*}

With respect to porosity, as demonstrated in \cref{fig:porosity}, the controlled cases show a clear improvement relative to BO, although porosity levels remain slightly higher than in the uncontrolled builds. 
This behavior is attributed to the controller’s symmetric penalization of parameter window violations, including those associated with insufficient energy input, which can promote localized lack-of-fusion defects. 
Nevertheless, the observed pore population in the controlled cases is characterized by reduced pore size and count compared to BO, suggesting a more favorable porosity distribution. 

Unlike previous approach~\cite{kavas2024situ} where controller tuning is performed to learn reference tracking for a single scan vector, our tuning procedure explicitly considers the full layer of an already overheating geometry while jointly accounting for tracking performance and input constraints. 
This leads to improved closed-loop behavior along the entire layer when the optimized controller is transferred to the real process.
The implemented tuning approach highlights the importance of component-specific controller design as an important future direction.

An important observation concerns the treatment of hard input constraints. Training the policy to adhere to constraints without explicitly bounding the inputs results in an unconstrained policy that rarely violates the constraints, providing additional flexibility in stabilizing the system dynamics. The porosity data in~\cref{fig:porosity}, however, indicates that such flexibility may cause problems to the microstructure, suggesting that hard thresholds should be enforced whenever possible, similar to the original control design. Conversely, the optimal input trajectory obtained in simulation without rate penalization, validated on hardware, shows that the closed-loop signal occasionally needs to exceed the predefined input limits. Therefore, the controller should consider both a time-varying output reference and input constraints for optimal performance. The interplay between process constraints and part microstructure is an interesting future direction of study.

\section{Conclusion} \label{Conclusion}

This paper proposes a multi-scale feedback control scheme using a dual-loop architecture integrating an in-layer and layer-to-layer controller to regulate the melt pool temperature in LPBF. 
The scheme is robust against unexpected uncertain parameters and computationally efficient. 
The controller tuned with the proposed method is implemented on a real LPBF machine, with specific adaptations made to accommodate the capabilities of the experimental setup, to showcase its effectiveness, highlighting the potential of boosting stabilization performance by employing a full dual-loop controller. 
Results show reliable performance, achieving superior geometric features when compared to state-of-the-art closed-loop control applications, with a minimal use of experimental data and an effective simulation-to-reality strategy.
Although oscillatory control inputs are consistently observed to reduce the vector-head transition and swelling, the physical mechanism remains unresolved and requires targeted experiments.
Future directions of the research include further improving the efficiency and accuracy of the layer-to-layer control loop and providing safety guarantees for the controller, as well as further studies on the robustness of the controller in various experimental settings.

\bibliographystyle{model1-num-names}
\bibliography{bibliography}

\appendix
\section{Definitions of mathematical notations}\label{sec:DefinitionNotations}


\begin{table}[H]
    \renewcommand{\arraystretch}{1.2} 
    \caption{Definitions of mathematical notations}
    \centering
    \begin{tabularx}{\columnwidth}{lX}
        \hline
        Notation & Description\\
        \hline
        $\rho$ & 3D position in metal powder\\
        $r, z$ & 2D cross-section position and height in metal powder corresponding to $\rho$\\
        $p$ & 2D cross-section trajectory of the laser beam defined by the scan strategy\\
        $T$ & Temperature\\
        $T_s, T_\infty$ & Build plate and ambient temperature\\
        $n$ & Number of nodes in each layer\\
        $n_{r_1}, n_{r_2}$ & Number of nodes on the first and second axis in each layer\\
        $\delta r_1, \delta r_2, \delta z$ & Length of each side of the node that occupies a rectangular volume\\
        $k, N$ & Layer index, $k$ denotes the $k$-th layer in $N$ layers\\
        $a$ & Cross-section radius of the laser beam\\
        $\kappa$ & Thermal conductivity\\
        $\kappa_\text{p}, \kappa_\text{d}, \kappa_\text{pd}$ & Thermal conductivity of powder, of solidified metal, and between powder and solidified metal\\
        $h_\infty$ & Convection coefficient between nodes and the atmosphere\\
        $\tau$ & Continuous time index\\
        $\tau_f, \tau_p, \tau_c$ & Processing time, printing time and recoating time for each layer\\
        $t, \Delta t$ & Discrete time index and sampling time interval\\
        $t_f, t_p, t_c$ & Number of sampling time interval corresponding to processing time, printing time and recoating time for each layer\\
        $c_h$ & Heat capacity\\
        $c_p, c_d$ & Heat capacity of powder and solidified metal. $c_p=(1-\epsilon)c_d$\\
        $y_N$ & Output for the whole layer $N$, formed by stacking output for each printing time step in layer $N$\\
        $\hat{y}_N$ & Measured output for layer $N$\\
        $y_d$ & Desired tracking output (scalar)\\
        $u_N$ & Input (laser power) for the whole layer $N$, formed by stacking input for each printing time step in layer $N$\\
        $u_N^f, u_N^b$ & Feedforward layer-to-layer control input, and feedback in-layer control input for layer $N$\\
        $\Pi_U$ & Projection onto the set U\\
        $X_N[0]$ & State for all nodes at time step 0 when printing layer $N$\\
        $Y^u_N, Y^x_N, Y^d_N, \bar{d}_N$ & System dynamics related matrices or vectors in the lifted system\\
        $L_y$ & Learning gain for norm-optimal ILC in layer-to-layer control\\
        $K$ & Feedback gain in in-layer control\\
        $\theta, \theta'$ & Parameter that influences the lifted system dynamics, and its corresponding estimate\\
        $\alpha$ & Fraction of the laser power absorbed by the material\\
        $\epsilon$ & Porosity of the metal\\
        $Q, R$ & Output and input regulating matrices for the layer-to-layer control loop\\
        \hline
    \end{tabularx}
    \label{Tab:MathematicalNotations}
\end{table}

\section{Model parameters for simulation and experiments}\label{sec:ModelParametersForSimulationAndExperiments}


\begin{table}[H]
    \renewcommand{\arraystretch}{1.2} 
    \caption{Model parameters for simulation}
    \centering
    \begin{tabularx}{\columnwidth}{lXp{3.5cm}}
        \hline
        Notation & Description & Value\\
        \hline
        $\delta r_1, \delta r_2$ & Length of each side of the node in cross-section & \num{2e-5}~[m]\\
        $\delta z$ & Height of the node & \num{5e-5}~[m]\\
        $n_{r_1}, n_{r_2}$ & Number of nodes on the first and second axis in each layer & 25\\
        $T_s$ & Build plate temperature & 900~[K]\\
        $T_\infty$ & Ambient temperature & 300~[K]\\
        $u_{\text{min}}$ & Minimum allowable value for laser power & 0~[W]\\
        $u_{\text{max}}$ & Maximum allowable value for laser power & 50~[W]\\
        $\tau_p, \tau_c$ & Printing time and recoating time for each layer & \num{1.25e-3}~[s]\\
        $\tau_f$ & Processing time for each layer & \num{2.5e-3}~[s]\\
        $\Delta t$ & Sampling time interval for time domain discretization & \num{1e-5}~[s]\\
        $Q$ & Output regulating matrix for the layer-to-layer control loop & 1000\\
        $R$ & Input regulating matrix for the layer-to-layer control loop & 1\\
        $h_\infty$ & Convection coefficient between nodes and the atmosphere & \num{10}~[W/(m\textsuperscript{2}\,K)]\\
        $\kappa_\text{p}$ & Thermal conductivity of powder & \num{0.5}~[W/(m\,K)]\\
        $\kappa_\text{d}$ & Thermal conductivity of solidified metal & \num{20}~[W/(m\,K)]\\
        $\kappa_\text{pd}$ & Thermal conductivity between powder and solidified metal & \num{10.25}~[W/(m\,K)]\\
        $c_d$ & Volumetric heat capacity of solidified metal & \num{4.25e6}~[J/(m\textsuperscript{3}\,K)]\\
        $\epsilon$ & Porosity of the metal & 0.6\\
        $\alpha$ & Fraction of the laser power absorbed by the material & 0.42\\
        $a$ & Cross-section radius of the laser beam & \num{6e-5}~[m]\\
        $L_y$ & Learning gain for norm-optimal ILC in layer-to-layer control & 0.8\\
        
        \hline
    \end{tabularx}
    \label{Tab:ProcessParameters}
\end{table}



\begin{table}[H]
    \renewcommand{\arraystretch}{1.2} 
    \caption{Model parameters for real machine}
    \centering
    \begin{tabularx}{\columnwidth}{lXp{3.5cm}}
        \hline
        Notation & Description & Value\\
        \hline
        $\delta r_1, \delta r_2$ & Length of each side of the node in cross-section & \num{3e-4}~[m]\\
        $\delta z$ & Height of the node & \num{3e-5}~[m]\\
        $n_{r_1}, n_{r_2}$ & Number of nodes on the first and second axis in each layer & 50\\
        $T_s$ & Build plate temperature & 900~[K]\\
        $T_\infty$ & Ambient temperature & 300~[K]\\
        $u_{\text{min}}$ & Minimum allowable value for laser power when laser is on & 140~[W]\\
        $u_{\text{max}}$ & Maximum allowable value for laser power when laser is on & 210~[W]\\
        $\tau_p, \tau_c$ & Printing time and recoating time for each layer & \num{1.2854e-1}~[s]\\
        $\tau_f$ & Processing time for each layer & \num{2.5708e-1}~[s]\\
        $\Delta t$ & Sampling time interval for time domain discretization & \num{1e-5}~[s]\\
        $h_\infty$ & Convection coefficient between nodes and the atmosphere & \num{10}~[W/(m\textsuperscript{2}\,K)]\\
        $c_d$ & Volumetric heat capacity of solidified metal & \num{4.25e6}~[J/(m\textsuperscript{3}\,K)]\\
        \hline
    \end{tabularx}
    \label{Tab:ProcessParametersReal}
\end{table}

\begin{table}[H]
    \renewcommand{\arraystretch}{1.2} 
    \caption{Model parameters derived from calibration}
    \centering
    \begin{tabularx}{\columnwidth}{lXp{3.5cm}}
        \hline
        Notation & Description & Value\\
        \hline
        $\kappa_\text{p}$ & Thermal conductivity of powder & \num{5}~[W/(m\,K)]\\
        $\kappa_{pd}$ & Thermal conductivity between powder and solidified metal & \num{1}~[W/(m\,K)]\\
        $\epsilon$ & Porosity of the metal & 0.5\\
        $\alpha$ & Fraction of the laser power absorbed by the material & 0.5\\
        $a$ & Cross-section radius of the laser beam & \num{9e-4}~[m]\\
        
        \hline
    \end{tabularx}
    \label{Tab:ProcessParametersRealCalibration}
\end{table}

\twocolumn

\end{document}